\documentclass[sigconf]{acmart}

\AtBeginDocument{%
  \providecommand\BibTeX{{%
    \normalfont B\kern-0.5em{\scshape i\kern-0.25em b}\kern-0.8em\TeX}}}

\newcommand \revision[1]{{\textcolor{black}{#1}}}

\setcopyright{acmcopyright}
\acmConference[CHI '23]{CHI' 23: The ACM CHI Conference on Human Factors in Computing Systems}{April 23 --- April 28, 2023 }{Hamburg, Germany}
\acmBooktitle{CHI '23: The ACM CHI Conference on Human Factors in Computing Systems, April 23 --- April 28, 2023, Hamburg, Germany}
\acmPrice{15.00}

\copyrightyear{2023} 
\acmYear{2023} 
\setcopyright{rightsretained} 
\acmConference[CHI '23]{Proceedings of the 2023 CHI Conference on Human Factors in Computing Systems}{April 23--28, 2023}{Hamburg, Germany}
\acmBooktitle{Proceedings of the 2023 CHI Conference on Human Factors in Computing Systems (CHI '23), April 23--28, 2023, Hamburg, Germany}\acmDOI{10.1145/3544548.3581008}
\acmISBN{978-1-4503-9421-5/23/04}

\begin{document}

\title[]{Enabling Voice-Accompanying Hand-to-Face Gesture Recognition with Cross-Device Sensing}

%% Towards Designing and Recognizing 
% Enabling Concomitant Hand-to-Face Gesture Sensing for Voice Interaction


\author{Zisu Li}
\authornote{indicates equal contribution.}
\authornote{This work was conducted when Zisu Li was a research intern at Tsinghua University.}
\email{zlihe@connect.ust.hk}
\orcid{0000-0001-8825-0191}
\affiliation{
  \institution{Tsinghua University}
  \country{Beijing, China}
}
\affiliation{
  \institution{The Hong Kong University of Science and Technology}
  \country{Hong Kong SAR, China}
}

\author{Chen Liang}
\authornotemark[1]
\email{liang-c19@mails.tsinghua.edu.cn}
\orcid{0000-0003-0579-2716}
\affiliation{
  \institution{Tsinghua University}
  \country{Beijing, China}
}

\author{Yuntao Wang}
\authornote{indicates the corresponding author.}
\email{yuntaowang@tsinghua.edu.cn}
\orcid{0000-0002-4249-8893}
\affiliation{
  \institution{Tsinghua University}
  \country{Beijing, China}
}

\author{Yue Qin}
\email{qiny19@mails.tsinghua.edu.cn}
\orcid{0000-0003-1351-5284}
\affiliation{
  \institution{Tsinghua University}
  \country{Beijing, China}
}

\author{Chun Yu}
\email{chunyu@tsinghua.edu.cn}
\orcid{0000-0003-2591-7993}
\affiliation{
  \institution{Tsinghua University}
  \country{Beijing, China}
}

\author{Yukang Yan}
\email{yukangy@andrew.cmu.edu}
\orcid{0000-0001-7515-3755}
\affiliation{
  \institution{Tsinghua University}
  \country{Beijing, China}
}

\author{Mingming Fan}
\email{mingmingfan@ust.hk}
\orcid{0000-0002-0356-4712}
\affiliation{
 % \institution{Computational Media and Arts Thrust}
  \institution{The Hong Kong University of Science and Technology (Guangzhou)}
  \city{Guangzhou}
  \country{China}
  }
\affiliation{
  \institution{The Hong Kong University of Science and Technology}
  \city{Hong Kong SAR}
  \country{China}
}

\author{Yuanchun Shi}
\email{shiyc@tsinghua.edu.cn}
\orcid{0000-0003-2273-6927}
\affiliation{
  \institution{Tsinghua University}
  \city{Beijing}
  \country{China}
}
\affiliation{
  \institution{Qinghai University}
  \city{Xining}
  \country{China}
}

\renewcommand{\shortauthors}{Li and Liang, et al.}
%, to indicate or convey certain intentions,

\begin{abstract}
Gestures performed accompanying the voice are essential for voice interaction to convey complementary semantics for interaction purposes such as wake-up state and input modality. In this paper, we investigated voice-accompanying hand-to-face (VAHF) gestures for voice interaction. We targeted on hand-to-face gestures because such gestures relate closely to speech and yield significant acoustic features (e.g., impeding voice propagation). We conducted a user study to explore the design space of VAHF gestures, where we first gathered candidate gestures and then applied a structural analysis to them in different dimensions (e.g., contact position and type), outputting a total of 8 VAHF gestures with good usability and least confusion. To facilitate VAHF gesture recognition, we proposed a novel cross-device sensing method that leverages heterogeneous channels (vocal, ultrasound, and IMU) of data from commodity devices (earbuds, watches, and rings). Our recognition model achieved an accuracy of 97.3\% for recognizing 3 gestures and 91.5\% for recognizing 8 gestures \revision{(excluding the "empty" gesture)}, proving the high applicability. Quantitative analysis also shed light on the recognition capability of each sensor channel and their different combinations. In the end, we illustrated the feasible use cases and their design principles to demonstrate the applicability of our system in various scenarios. 

\end{abstract}







\begin{teaserfigure}
  \centering
  \includegraphics[width=0.8\textwidth]{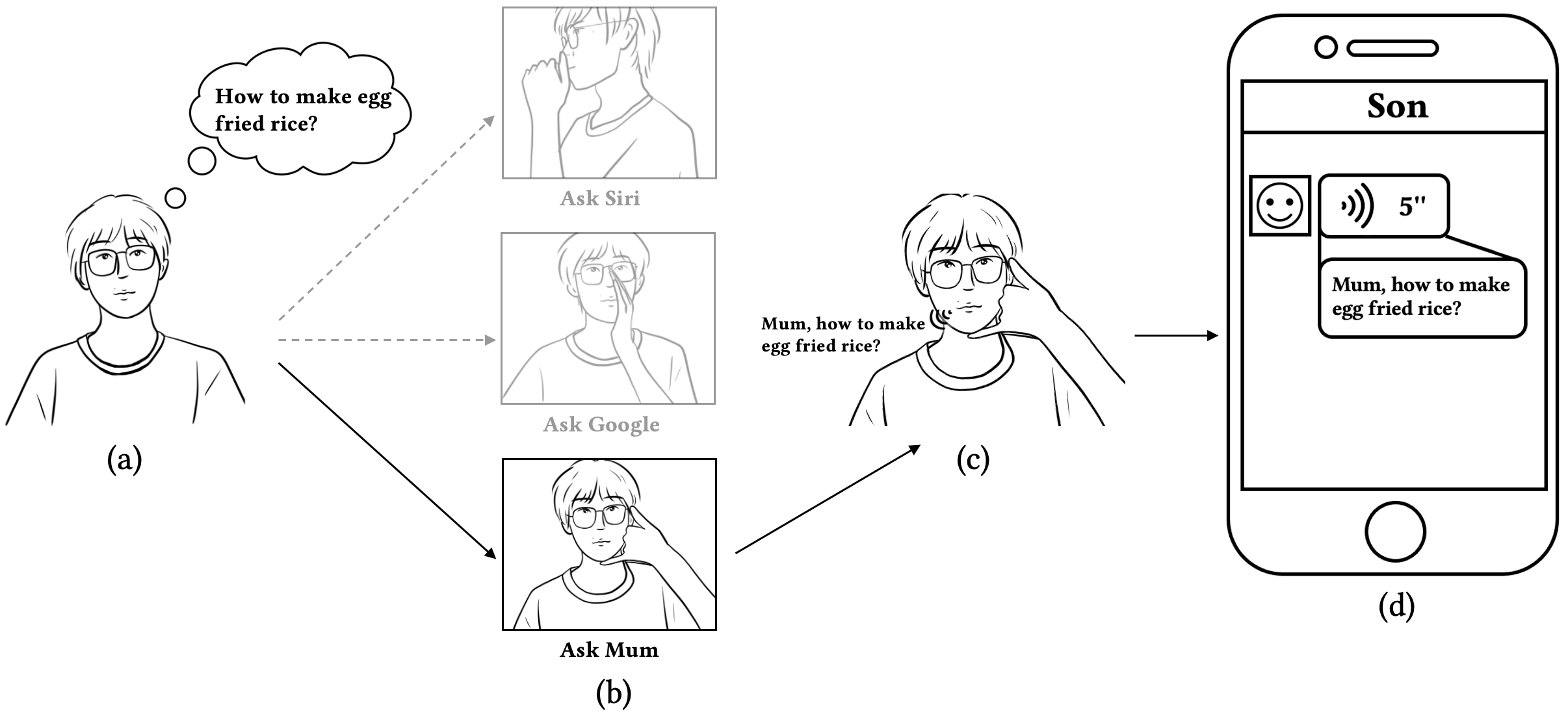}
  \caption{A typical usage scenario enabled by voice-accompanying hand-to-face (VAHF) gestures. (a) The user wants to know how to make egg fried rice. (b) The user can perform different VAHF gestures to redirect their voice input to different targets (e.g., asking Siri, searching on Google with the transcribed text, and sending a voice message to mum). (c) The user performs a "phone call" gesture and speaks simultaneously. (d) The smart devices recognize the user's intention through the performed VAHF gesture and simulate sending a voice message to the user's mum.}
  \Description{This figure contains a typical usage scenario enabled by voice-accompanying hand-to-face (VAHF) gestures. (a) The user wants to know how to make egg-fried rice. (b) The user can perform different VAHF gestures to redirect their voice input to different targets (e.g., asking Siri, searching on Google with the transcribed text, and sending a voice message to mum). (c) The user performs a "phone call" gesture and speaks simultaneously. (d) The smart devices recognize the user's intention through the performed VAHF gesture and simulate sending a voice message to the user's mum. }
  \label{fig:teaser}
\end{teaserfigure}

\maketitle
\section{Introduction}
% 1. Voice interaction for wearable devices is common, e.g., earphones. However, interactions solely rely on voice is not convenient with two major reasons. 1) requirement of the wake-up keyword. 2) Function navigation in a voice menu requires multiple rounds of communications. 

% 2. Adding gesture to the voice input is a promising approach to rich the functionality. 

Voice input has become a natural and always-available interaction modality for wearable devices such as earphones and smartwatches. However, the modality control (e.g., the wake-up state) in voice interaction is still a challenging problem due to the implicitness of modality information in speech and the restricted NLP techniques. People have to repeat the hotword to switch to the modality or the target device actively, which introduce extra burdens for the interaction. Thus, researchers have been seeking supplementary input methods as parallel input channels to assist voice interaction \revision{\cite{10.1145/3351276,Yan-UIST-2019,10.1145/3313831.3376479}}.

%(e.g., PrivateTalk \cite{Yan-UIST-2019}).
% However, interactions solely rely on voice is not convenient for the following two primary reasons. Wake-up keywords were required to activate the voice interface. Navigation in a voice menu requires multiple rounds of voice communications. Solving these issues can simplify the interaction flow and improve the user experience. Therefore, researchers are actively seeking supplementary solutions to facilitating voice interaction for wearable devices. 

% \revision{In the aspect of interaction efficiency, the combination of voice and gesture input can overcome the typical computer interaction trade-off: ease V.S. expressiveness} \revision{ since speech and gestural commands are easy to execute while retaining a large command vocabulary \cite{10.1145/307710.307730}}. \revision{In the aspect of linguistics }, p
People tend to perform body gestures accompanying their voice for better expression of certain emotions or intentions during the conversation \revision{\cite{10.1007/978-3-030-05792-3_15}}. In such a case, we defined the gesture, which is performed simultaneously with the speech, as a voice-accompanying gesture. Analogously, in the voice interaction with smart devices, the use of voice-accompanying gestures could provide parallel information that expands the input channel bandwidth \revision{\cite{Katsamanis2017,10.1145/3313831.3376479,VoGe2017}}, simplifies the voice interface flow \revision{\cite{10.1145/3351276,10.1145/3411764.3445687,Yan-UIST-2019}}, and make voice interaction more convenient \revision{\cite{10.1145/3411764.3445687,Yan-UIST-2019,10.1145/3313831.3376479}}. \revision{The underlying logic why voice-accompanying gestures have been prevalent and widely researched is due to the physiological nature that the voice channel and the gesture channel are highly independent and complementary \cite{10.1145/3313831.3376479} (e.g., performing a gesture as parallel input information while not repressing the voice expressivity).} For instance, the user can define a specific voice-accompanying gesture (e.g., covering the mouth) instead of repetitively saying the wake-up keyword to keep the voice interface active. The user can also define multiple gestures to represent the redirection of voice input to different input modalities (e.g., ignored, interrupted, transcribed, or raw audio input), target devices (e.g., whether should the TV or the smartphone accept the input), and shortcuts (e.g., binding certain UI operation flows with the gesture).

% todo: R1 suggests enhancing the introduction by emphasizing what makes voice accompanying hand-to-face gestures more suitable than other input methods.
% other methods: raise-to-speak, proxitalk, xxx
% VAHF 的好处 - voice accompanying - parallel input channel, hand-to-face - natural, related to voice
%\revision{Why we chose hand-to-face gesture motivated by XXX proven to be natural and more related to the speec}

In this paper, we investigated the feasibility of using voice-accompanying hand-to-face (VAHF) gestures as parallel channels to improve the traditional voice interaction flow. Specifically, we aim at designing VAHF gestures and recognizing them with an acoustic-based cross-device sensing method. We targeted hand-to-face gestures as the instantiation of voice-accompanying gestures because they \revision{have been proven to be natural, expressive (e.g., various landmarks on the face to yield a large gesture space), and more related to the speech by existing researches \cite{earbuddy,Yan-UIST-2019,46-facesight}}. Moreover, hand-to-face gestures yielded significant features in voice propagation, which is beneficial for acoustic sensing.

% In this paper, we aim at designing para-linguistic hand-to-face (PLHF) gestures and recognizing them with a novel cross-device sensing method.

To understand the design space of VAHF gestures, we conducted a user-centric gesture elicitation study with a total proposal of 15 gestures from end-users. Then we narrowed down the gesture set from 15 to 8 gestures with better usability and the least ambiguity. 
% not only usability and ambiguity, but also sematically related

As for the sensing schemes, the underlying principle is that when the user speaks, each VAHF gesture creates a unique acoustic propagation path from the mouth to the set of microphones on wearable devices. Therefore, our method can recognize the hand-to-face gesture using the acoustic features of each unique propagation path. Further, we incorporated an ultrasound channel and an IMU channel to provide supplementary sensing information and enhance gesture recognition. For the ultrasound channel, the smartwatch served as an ultrasound source and all microphones captured such signals from different positions to indicate position-aware features. For the IMU channel, the IMU on the ring can convey the attitude and motion of the user's hand. We also investigated the fusion mechanism among different devices and channels.

To evaluate our technique, we first built a cross-device VAHF dataset consisting of 1800 samples of 8 VAHF gestures \revision{along with one "empty" gesture (meaning not performing a VAHF gesture)}. Then we conducted 1) a two-factorial evaluation regarding sensor combination and model selection, 2) an extensive evaluation of a reduced gesture set, and 3) an ablation study for the optimal model to validate the computation feasibility and the applicability of our technique. Results showed our model achieved high recognition accuracy of 97.3\% for 3 + 1\revision{(empty)} gestures and 91.5\% for 8+1\revision{(empty)} gestures recognition on our cross-device VAHF dataset. Quantitative analysis also sheds light on the recognition capability of each sensor channel and its different combinations.

At the end of the paper, we discuss real-life application scenarios to demonstrate the applicability of VAHF gestures as well as provide general design implications for VAHF gesture-enhanced voice interaction.

In summary, the contributions of this paper are as follows:
\begin{itemize}
    \item We conducted a comprehensive study to elicit the gesture space of VAHF gestures and proposed a gesture set with better usability, better social acceptance, less fatigue, and less ambiguity. 
    
    \item We propose a novel sensor-fusion technique for VAHF gesture recognition which is supported by cross-device sensors. Our quantitative analysis sheds light on the recognition capability of the different sensor combinations over VAHF gestures with different characteristics.
    \item We demonstrate a set of use cases of our gesture recognition technique that outline new opportunities for VAHF gestures to benefit voice interaction.
\end{itemize}

\section{Related Work}
In this section, we presented related work in three aspects: enhancing voice interaction with parallel gestures, hand-to-face interaction, and cross-device sensing for hand gestures.

%First, we review previous research to enhance voice interaction with parallel gestures. We then summarize the combination of the use of voice and hand gesture. Finally, we introduce different sensing technologies for \projectName{} gestures. 

%Gestures as pl channel to enhance voice interaction
%PL gesture design
%cross device sensing for hand gestures

% Para-language, which includes gestures, facial expressions, interfactional synchrony, eye contact, use of space, touching, aspects of voice modification, and silence are shown to play a crucial role in human interaction \cite{pennycook1985actions}.

\subsection{Enhancing Voice Interaction with Parallel Gestures}
Performing body gestures parallel to voice commands has been a prevalent method to convey certain intentions or information during the voice interaction. People tended to use gestures of different body segments, such as head gestures\cite{1374748,9053124,10.1145/1088463.1088470,10.1145/3313831.3376479}, gaze\cite{4115628, 10.1145/2043674.2043723}, hand gestures \cite{Yan-UIST-2019}, and facial expressions \cite{10.1145/3313831.3376810}, to provide supplementary information obliged for voice interaction. The purpose of introducing of certain parallel body gestures to the voice interface typically included: indicating the wakeup state \cite{10.1145/3411764.3445687,Yan-UIST-2019,10.1145/3351276}, serving as the control (or trigger) signal \cite{earbuddy,10.1145/3313831.3376810}, and passing scene-related context information \cite{10.1145/3313831.3376479,10.1145/3379337.3415588}. For example, Yan et al. \cite{10.1145/3313831.3376810} proposed frowning, a facial expression of para-language, to implement interrupting the responses during voice interactions between human and smart devices. Qin et al. \cite{10.1145/3411764.3445687} leveraged the speech features when user raise the microphone embeded in devices to close to mouth to facilitate wake-up free techniques. WorldGaze \cite{10.1145/3313831.3376479} used commodity a smartphone to recognize the real-world head-gaze location (e.g., certain buildings or objects) of a user to provide the voice agent with supplementary scene-related information for better comprehension.
% Researchers have proved the importance and benefits when introducing para-language to enhance voice interaction. For example, Nomoto et al. \cite{nomoto2011anger} applied para-linguistic cues named dialog-features such as turn-taking and backchannel feedback to recognize anger in conversational interactive situations more robustly. Pennycook et al. \cite{pennycook1985actions} examined the importance of para-language (kinesics, proxemics, and paraverbal features) in communication and proposed some suggestions to facilitate students' acquisition of para-language. 

% In addition to leveraging para-language to enhance the machines' understanding of semantics or user behaviour in voice interaction, researchers also attempted to use para-language as active input with the aim of realizing a faster and more natural way of voice interaction. For example, Yan et al. \cite{10.1145/3313831.3376810} proposed frowning, a facial expression of para-language, to implement interrupting the responses during voice interactions between human and smart devices. Qin et al. \cite{10.1145/3411764.3445687} leveraged the speech features when user raise the microphone embeded in devices to close to mouth to facilitate wake-up free techniques. Yan et al. \cite{Yan-UIST-2019} introduced an interaction technique which allow users to activate voice input by performing the Hand-On-Mouth gesture during speaking. 

Our work was also within the framework of using parallel gestures as active input to enhance voice interaction, which was most related to but achieved a leap over PrivateTalk \cite{Yan-UIST-2019}, which allowed users to activate voice input by performing a hand-on-mouth gesture during speaking. Compared with existing work \cite{10.1145/3313831.3376810,10.1145/3411764.3445687,Yan-UIST-2019} where a specific gesture (e.g., bringing the phone to the mouth\cite{10.1145/3411764.3445687}) was designed and recognized for specific functionality (e.g., interrupting the conversation or activating the voice assistance), our multiple VAHF gesture recognition on multi-modal wearable devices could effectively broaden the input channel of actions as parallel information with the potential of supporting a larger interaction space, such as defining multiple shortcuts.

\subsection{Hand-to-Face Interaction}
%validation/priority
%design space
Gestures involving hand and face have been demonstrated as a natural and easy-to-use way to input commands. Prior research has proposed the validation of the inherent unobtrusiveness, subtlety and social acceptability\cite{HWDs} of hand-to-face interaction. Design space of hand-to-face gestures has been explored by prior research. For example, different face regions such as the ear \cite{EarTouch}, cheek \cite{cheek,HWDs} or nose \cite{nose} were demonstrated to have the viability of hand-to-face input. Mahmoud et al. \cite{hand-over-gesture_book} proves that people prefer lower face regions to upper regions, especially chin, mouth and lower cheeks, for naturalistic interaction. Weng et al. \cite{46-facesight} proposed recognition of hand-to-face gestures for AR glasses. Serrano et al. \cite{HWDs} provided a set of guidelines for developing effective Hand-to-Face interaction techniques and found that the cheek is the most promising area on the face. Miniaturizing obfuscating, screening, camouflaging and re-purposing have been purposed by Lee et al.\cite{Acceptable} as the strategies of the design of socially acceptable hand-to-face gestures. We refer to some principles (e.g. lower face region, social acceptance, etc.) from previous work mentioned above to design our subjective evaluation process. Different from previous work, our design space of hand-to-face gestures are generated from real-life conversations, which has more para-linguistic features resulting in performing more easily and naturally. 

Combining voice input with hand gestures can provide rich information to enable convenient and expressive multi-modal interaction~\cite{Katsamanis2017}. "Put that there" was the first multi-model interaction method introducing the pointing gesture to indicate the location mentioned in the voice command~\cite{Bolt1980}. Bourguet et al. studied the temporal synchronization between speech (Japanese) and hand pointing gestures during multi-modal interaction~\cite{Bourguet1998}. Sauras-Perez et al. \cite{VoGe2017} proposed a human vehicle interaction system based on voice and pointing gestures that enables the user making spontaneous decisions over the route and communicate them to the car. Closest to our work, Yan et al. \cite{Yan-UIST-2019} proposed to enable hand-to-mouth gesture interaction as a wake-up action for voice interfaces. However, the combination of the hand-to-face gestures and voice interaction in previous work is limited to few types of gestures and the fixed usage patterns. In our work, we discussed more types of gestures which can be used in versatile \revision{scenarios} of voice interaction.

\subsection{Cross-Device Sensing for Hand Gestures}
%先单音频、单imu，再写cross-device，还是一起写？
Free-form hand gesture sensing is key to enabling rich interaction space taking advantage of the expressiveness of human hand. Previous literature has investigated sensing solutions for free-form hand gestures with different sensors including cameras (RGB \cite{17-interface,19-fish-eye,30-Generate-hands-tracking,55-handsee}, IR \cite{3-IR-touch-on-phone,46-facesight}, and depth \cite{1-depth,2-depth}), EMG sensors \cite{1-EMG,2-EMG}, capacitive sensors \cite{1-capacitive}, millimeter-wave radar \cite{10.1145/2897824.2925953,10.1145/2984511.2984565,7907299}, acoustic sensors \cite{Wang2019,gupta2012soundwave,Ono2013,Liu2019}, and inertial sensors \cite{7-arm,31-ViBand,59-Smartwatch-Based,62-gesture-with-acc}. Among these sensors, the camera is most widely researched since it has the strongest sensing capability to capture pixel-wise image data, based on which many computer vision models have been developed for fine-grained hand sensing such as detecting hand keypoints and recovering the hand pose. However, vision-based hand gesture sensing often requires externally-mounted camera and heavy computation (e.g., using a GPU), which prevents the practical use in mobile and pervasive scenarios. Similarly, sensing methods based on EMG sensors \cite{1-EMG,2-EMG}, capacitive sensors \cite{1-capacitive}, and millimeter-wave radar \cite{10.1145/2897824.2925953,10.1145/2984511.2984565,7907299} require additional wearing on the human body, making them far from practical deployment. In our work, we focused on the latter two - microphones and inertial sensors - because they are computational efficient and largely equipped on commodity devices such as smartphones, earbuds, smartwatches, and smart rings. Below we presented related work on acoustic- and inertial-based hand gesture sensing. 

\subsubsection{Acoustic sensing}

The principle of acoustic sensing for hand gestures is to measure how specific hand gesture influences the propagation of active (e.g., active ultrasound) or passive (e.g., human voice) sound sources or makes a sound. Based on the presence of an active sound source, it can be categorized into active acoustic sensing and passive acoustic sensing.

Active acoustic sensing methods~\cite{Wang2019} has been widely explored for gesture recognition including in-air hand gesture~\cite{Yuzhou-reflectrack, gupta2012soundwave}, face orientation~\cite{yuntao-faceori}, ambient activity~\cite{Mesaros2010}, touch gesture on everyday objects or surfaces~\cite{Ono2013, Liu2019}, finger tracking~\cite{Yun2017}, silent speech interface~\cite{Gao2020, Zhang2020}, cough monitoring~\cite{wang-hearcough} or sleep activity recognition~\cite{Ken-sleep}. For example, Touch \& Activate~\cite{Ono2013} enabled touch interface on everyday objects by measuring the acoustic frequency response of the object and the touch gesture.
Strata~\cite{Yun2017} enabled 2-dimensional finger position tracking using the reflective impulse audio signal on commodity smartphones. Ando et al. \cite{Ando2017} modeled the transfer function, or the propagation path of the sound, and used it in gesture recognition. These methods recognized the acoustic echo or propagation features caused by the gestures using a speaker and one or more microphones. Doppler effect was also frequently used for sensing subtle hand gestures involving relative movements~\cite{gupta2012soundwave,Liu2019}. EchoWhisper ~\cite{Gao2020} also leveraged the Doppler shift of reflection for near-ultrasound sound waves caused by the mouth and tongue movements to interpret the speech and build a silent speech interface.

Passive acoustic sensing recognizes sound activities using merely microphones~\cite{Wu2020, Laput2018,Harrison2012,Harrison2011,Xiao-2014-Toffee, Mesaros2010}. For instance, Toffee~\cite{Xiao-2014-Toffee} enabled an ad-hoc touch interface on a table around the device using time of arrival correction. TapSense~\cite{Harrison2011} enhanced finger interaction on touch screens by detecting the unique sound features of fingertip, pad, knuckle, and nail. Acoustic Barcodes~\cite{Harrison2012} was an identifying tag that used notches to produce sound when dragged across, which can be recognized by a microphone for information retrieval or triggering interactive functions. Daily activities or ambient environment (e.g., taking a bus) can be detected based on features of the sound collected by a single microphone~\cite{Mesaros2010} or a microphone array. Ubicoustics \cite{Laput2018} proposed a plug-and-use sound recognition pipeline for general-purpose activity recognition. Wu et al.~\cite{Wu2020} further extended the environment acoustic event detection using an end-to-end system for self-supervised learning of events labeled through one-shot interaction.

In our work, we took advantage of both passive and active acoustic sensing. To be more specific, passive acoustic sensing recognized the hand's influence on the features of the accompanying speech, including frequency response, amplitude, etc. Active acoustic sensing helped to determine relative position-based features among different devices. The two channels can provide supplementary capability in VAHF gesture sensing.

\subsubsection{Inertial Sensing}
Inertial sensors, commonly integrated into commodity devices, are efficient in detecting motion- and attitude-related hand/finger gestures. For instance, a number of previous works \cite{7-arm,31-ViBand,59-Smartwatch-Based,62-gesture-with-acc,10.1145/3478114,10.1145/3569463} used acceleration and rotation with wrist-worn inertial sensors to recognize hand gesture. Serendipity \cite{Serendipity} recognized five fine-grained gestures based on the IMU in off-the-shelf smartwatches. Mo-Bi \cite{Mo-Bi2016}  used a smartphone and two accelerometer-embedded wrist-worn devices for each hand to collect the hand-posture data and developed the implicit hand-posture recognition software. Leveraging inertial sensors integrated into smartwatches, Float \cite{SunkeFloat2017} recognized wrist-to-finger gestures to enhance one-hand and smartwatch interaction. Gu et al. \cite{Gu-IMU-RING-TYPING} enabled one-finger typing with an index-finger-worn IMU ring by detecting hand-to-surface touching events and rotation angles. Lu et al. \cite{Lu-handtohand} studied the sensing capability of dual wrist-worn devices and analyzed cross-device features for more accurate gesture inference. 

Inspired by these works, we incorporated an IMU ring into our sensing system to capture the motion features of VAHF gestures. 

\subsubsection{Sensor Fusion Methods for Hand Gestures}
Previous research has explored cross-device sensor fusion methods to enhance 
the recognition capability of different types of hand gestures, especially when the sensing capability of different sensors are complementary for recognizing different features. Sensor fusion methods included homogeneous fusion and heterogeneous fusion. Homogeneous fusion aimed to add more homogeneous sensor nodes into the sensing system to capture fine-grained information. For example, fusing camera captures from different views \cite{moon2020interhand2} is a classical and effective solution to reduce 3D reconstruction or detection error which is also widely used in generating high-quality machine annotations. For acoustic sensing, adding more microphones to the scene achieved more fine-grained acoustic measurements, which is beneficial to various sensing purposes such as 2D localization \cite{Yun2017} and gesture classification \cite{Wang2019}.
%特别是当sensors对于传感某类手势的不同特征的能力互补的时候

The other type is heterogeneous fusion, where different types of sensors are combined to merge their strengths \cite{UWB-and-Doppler,EMG-and-camera,RFID,EMG-and-IMU-for-stroke,Kinect-and-IMU}. For example, Li et al. \cite{UWB-and-Doppler} presented a hierarchical sensor fusion approach for human micro-gesture recognition by combining an Ultra Wide Band (UWB) Doppler radar and wearable pressure sensors. Ceolini et al. \cite{EMG-and-camera} presented a sensor fusion framework that integrates complementary systems: the electromyography (EMG) signal from muscles and visual information. Ceolini et al. \cite{EMG-and-camera2019} also investigated the fusion of EMG and a camera on limited computational resources of mobile phones to detect gestures. A more typical scene is fusing an IMU with a camera \cite{10.1145/3332165.3347947}, where the IMU detects the subtle contact signal and the camera senses the global hand state. Acoustico \cite{10.1145/3379337.3415901} fused acoustic and IMU signal for 2D tap position localization based on the TDOA of the tap sound's two propagation paths.

In our work, we adopted both homogeneous fusion and heterogeneous fusion strategies. The former aimed to probe more measurement nodes into the sensing space, while the latter aimed to capture different types of features from different channels.

\section{Designing Voice-Accompanying Hand-to-Face Gestures}
To thoroughly understand and explore the gesture space of voice-accompanying gestures, we conducted a user-centric gesture elicitation study to elicit gesture design from end users. Subsequently, we conducted a hierarchical analysis process to narrow down the gesture space from 15 gestures to 8 gestures which are easy to perform, easy to memorize, and with less ambiguity.  \revision{The design of this study was in line with the previous gesture elicitation work \cite{10.1145/3461778.3462004,Lu-handtohand,earbuddy} consisting the typical phases of gesture proposal, gesture evaluation, and gesture set refinement.}

%Therefore, we generated empirical categories and collected users' subjective perception in different dimensions on the chosen gestures to identify a subset of the most preferable gestures. 

%我们通过brainstorming的方式先选出了一个general hand-to-mouth gesture set，之后通过经验性分类和主观测评最终从15个gesture中筛选出8个gestures。

\begin{table}
  \vspace{-0.3cm}
  \centering
  \caption{Text description and empirical categorization for all the 15 gestures in our gesture set. Each gesture was empirically categorized in three dimensions: contact position, contact type, and occlusion state. Contact position is represented in 5 levels: ear (E), mouth (M), chin(CN), cheek(CK), and none(N). Contact type is represented in 3 levels: finger(F), palm(P), and hand segments(HS). The occlusion state on the sound propagation path for the human voice to ears is represented in 3 levels: hardly(H), partially(P), and completely(C).  }
  \Description{This table contains text description and empirical categorization for all the 15 gestures in our gesture set. Each gesture was empirically categorized in three dimensions: contact position, contact type, and occlusion state. Contact position is represented in 5 levels: ear (E), mouth (M), chin(CN), cheek(CK), and none(N). Contact type is represented in 3 levels: finger(F), palm(P), and hand segments(HS). The occlusion state on the sound propagation path for the human voice to ears is represented in 3 levels: hardly(H), partially(P), and completely(C). }
  ~\label{tab:gestures}
    \vspace{-0.3cm}
    \resizebox{1\columnwidth}{!}{
    \begin{tabular}{l|c|c|c|c|c}
    \toprule
    Index & Gesture & Contact Position & Contact Type & Occlusion State & \revision{Semantics} \\
    \midrule
    1  & pinch ear rim & E &  F & H & \revision{Earphone Manipulation}\\
    \midrule
    2  & thinking face gesture & M, CN & HS & H & \revision{Thinking, Querying}\\
    \midrule
    3  & support cheek with fist  &  M, CK & P & P & \revision{Thinking, Resting}\\
    \midrule
    4 & non-contact cover mouth with palm   &  N &	P &	P & \revision{Directional Speech, Whisper}\\
    \midrule
    5 & support cheek with palm  &  M, CK & P & P & \revision{Thinking, Concentrating}\\
    \midrule
    6  & cover mouth with fist  & M & HS & C & \revision{Interphone, Messaging}\\
    \midrule
    7  & cover ear with arched palm & E & P & C & \revision{Hearing, Phone Call}\\
    \midrule
    8  & hold up palm beside nose and mouth & M, CK& P& C & \revision{Directional Speech, Block}\\
    \midrule
    9  & touch earphone with index finger &  E &  F & H & \revision{Earphone Manipulation}\\
    \midrule
    10  & touch top ear rim  &  E &  F & H & \revision{Earphone Manipulation}\\
    \midrule
    11  & touch vocal cord   &  N&F& H & \revision{Voice Distortion}\\
    \midrule
    12  & cover mouth with palm & M, CK, CN & P & C & \revision{Silence, Whisper}\\
    \midrule
    13  & shushing gesture   & M& F& H & \revision{Silence, Interruption}\\
    \midrule
    14  & touch the back of ear rim &  E &  F & H & \revision{Hearing, Attention}\\
    \midrule
    15  & calling gesture &  M, CK, CN, E& HS & P & \revision{Communication, Phone Call}\\
    \bottomrule
  \end{tabular}
  }
  \vspace{0.2cm}
\end{table}

\subsection{Voice-accompanying Hand-to-Face Gesture Proposal}
We conducted a brainstorming gesture proposal study to understand users' preference on VAHF gestures and derive a gesture set with general agreements. 

\subsubsection{Participants, Brainstorming Design, and Procedure}
%participant
We recruited 10 participants (4 female, all right-handed) from the local campus, with an average age of 21.3 (from 18 to 27, SD=2.4). Their familiarity score of wearable devices and voice interaction was 3.3\/5 (SD=1.2). The whole study took about 1 hour and each participant received 15\$ for compensation. 

%design
The purpose of this study is to encourage participants to brainstorm as many voice-accompanying hand-to-face (VAHF) gestures as they could without considering the sensing feasibility \revision{and together work out a usable VAHF gesture set with common agreements}. To achieve this point, we do not restrict the gestures to specific application scenarios or tasks, which maintained their focus on the physical nature of performing different VAHF gestures. The only constraint we imposed on the design was that the gestures should be static and durable to meet the nature of "voice-accompanying". The whole study consisted of 4 stages: 1) icebreaking and introduction, 2) individual thinking, 3) individual proposal, and 4) group discussion.

%  The only constraint we imposed on the design was that the gestures should imply para-linguistic semantics or have been used in current voice interaction.

%procedure
After a short icebreaking procedure where all the participants introduced themselves and familiarized themselves with each other, the experimenter acknowledged the participants of the purpose and the procedure for the study as well as the definition of VAHF gestures to the participants. Then the participants went through an individual thinking process for 10 minutes where participants worked separately to \revision{come} up with as many gestures as possible and wrote them down on a notebook \revision{as detailed as possible (e.g., encouraging them to write down the motivations, semantics, and potential usages of the gestures other than simply the descriptions).} After that, each participant was asked to verbalize their proposal \revision{(including the gesture descriptions, motivations, semantics, and potential usages)} and perform the proposed gestures by hand orderly. They could also sketch and show their ideas on a public whiteboard. Participants then came up with a group discussion where one could either show the pros and cons of the others' proposal or generate new gestures from the others' inspiration. The discussion ended until all participants worked out a final gesture set with the consistent agreement. \revision{The whole brainstorming process was hosted by two experimenters - one guiding the experiment while the other taking notes of the key points presented by the participants.}

\subsubsection{Results and Discussion}
\begin{figure*}
    \centering
    \includegraphics[width=1\textwidth]{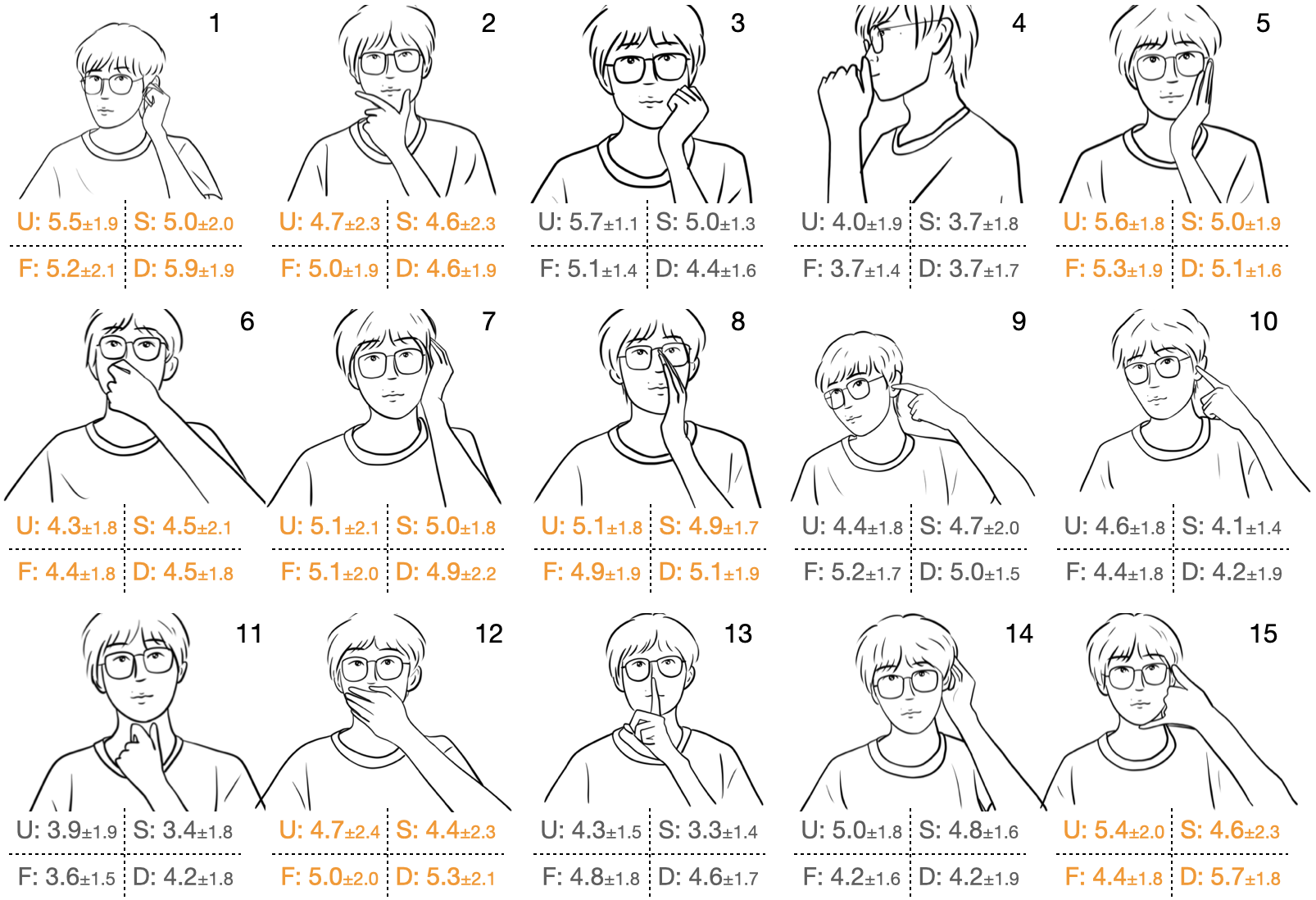}
    \caption{Drafts illustrating each gesture in the gesture set: 1) pinch the ear rim, 2) thinking face gesture, 3) support cheek with fist, 4)non-contact cover mouth with palm, 5)support cheek with palm, 6) cover mouth with fist 7) cover ear with arched palm, 8) hold up palm beside nose and mouth,  9) touch the earphone with index finger, 10) touch the top ear rim, 11) touch the vocal cord, 12) cover mouth with palm, 13) shushing gesture, 14) touch back of ear rim with fingers, 15) calling gesture. The mean±s.d. of users’ subjective scores (1-7, the higher the better) on Usability (U), Social Acceptance (S), Fatigue (F) and Disambiguity (D) is shown on the bottom of each draft. Scores of gestures in our final gesture set are highlighted in orange. }
    \Description{This figure contains each gesture in the gesture set: 1) pinch the ear rim, 2) thinking face gesture, 3) support cheek with fist, 4)non-contact cover mouth with palm, 5)support cheek with palm, 6) cover mouth with fist 7) cover ear with arched palm, 8) hold up palm beside nose and mouth,  9) touch the earphone with index finger, 10) touch the top ear rim, 11) touch the vocal cord, 12) cover mouth with palm, 13) shushing gesture, 14) touch back of ear rim with fingers, 15) calling gesture. The mean±s.d. of users’ subjective scores (1-7, the higher, the better) on Usability (U), Social Acceptance (S), Fatigue (F) and Disambiguity (D) is shown on the bottom of each draft. Scores of gestures in our final gesture set are highlighted in orange. }
    \label{fig:gestures}
\end{figure*}

Fig. \ref{fig:gestures} and the "Gesture" column of Table \ref{tab:gestures} illustrated the 15 VAHF gestures and their text descriptions proposed by users in the brainstorming study. \revision{The "Semantics" column summarized some of the typical semantics of each gesture from participants' quotes. An interesting finding is that participants tended to design the gestures in a mimetic and semantic-based manner, borrowing the inspirations from their daily activities and usages of smart devices. For example, touching gestures on the ear (gestures 1, 9, and 10 in Table \ref{tab:gestures}, same as below) was regarded as the metaphor of earphone manipulations related to voice interaction (N=9), which came from their experiences of using wireless earbuds (e.g., triggering the voice assistant and controlling the volume and the progress). Participants also presented their potential usages, such as waking up Siri (P2), making a voice memo (P3), and sending a voice message to a specific person (P8). Similarly, participants described gestures 6, 7, and 15 as "the imitation of using certain devices" (N=10). \textit{"Holding up the fist in front of the mouth is a cool gesture. It is just like sending an instant command with an interphone"} (P4). \textit{"Covering the ear with the arched palm is like you are holding the phone while the 'calling' gesture is like you are imitating an old-fashioned telephone. I would prefer the former one because it is easy to perform and seems more natural to others"} (P10).}

\revision{In addition to the above gestures related to device usage, some gestures were proposed for their prevalence in daily communication and social expression. Participants (N=10) showed their will to transfer these gestures to the interaction with voice assistance. For example, the "shushing" gesture (gesture 13) and the "flaring ear" gesture (gesture 14) were proposed because they were frequently used in daily dialog. \textit{"Shushing has the meaning of silence and interruption. We can also use it to interrupt the conversation with the voice assistant" (P3).} \textit{"The 'flaring ear' gesture means 'pardon' or shows attention to the speaker. I guess it would be nice to assign this gesture to functions with similar meanings" (P1).} Gestures 4, 6, 7, and 12 were mouth-related gestures proposed by the participants, with the general meanings of special speech, lowered volume, whisper, and silence. The gestures were distinguished by different ways of covering the mouth. \textit{"When I hold up the palm on one side of the mouth, I probably want to speak to the one on the other side directionally. However, when I cover my mouth, the meaning could be totally different" (P3).} Similarly, participants designed three face-related gestures (gestures 2, 3, and 5), indicating thinking, querying, resting, or concentrating, yet with slightly different implications. \textit{"It would be wonderful the voice assistant could respond to my 'thinking face' gesture by querying my words on the searching engine"} (P2). Exceptionally, P5 proposed a "touch vocal cord" gesture (gesture 11) with a unique position. "The vocal cord affects the timbre, meaning to 'make a different sound' (P5)."}

\revision{Although the semantics of each gesture seemed clear to individuals, we found some conflicts in the group discussion stage. For example, regarding the "cover mouth with fist" gesture (gesture 6), most participants showed approval of the "interphone" metaphor while some participants (P2, P5) thought it should be with the semantics of "silence" and "secrete". Some participants also mentioned the meanings and preferences of certain gestures might vary under different cultural backgrounds, especially for the gestures with social functionalities.}

% 1	pinch the ear rim	E	F	H
% 2	thinking face gesture	M, CN	HS	H
% 3	support cheek with fist	M, CK	HS	P
% 4	non-contact cover mouth with palm	N	P	P
% 5	support cheek with palm	M, CK	P	P
% 6	cover mouth with fist	M	HS	C
% 7	cover ear with palm	E	P	C
% 8	cover nose and mouth with palm	M, CK	P	P
% 9	touch the earphone with index finger	E	F	H
% 10	touch the top ear rim	E	F	H
% 11	touch the vocal cord	N	F	H
% 12	cover mouth with palm	M, CK, CN	P	C
% 13	shushing gesture	M	F	H
% 14	touch the back of ear rim with finger tip	E	F	H
% 15	calling gesture	M, CK, CN, E	HS	P

\subsection{Optimizing VAHF Gesture Set}
To derive the final user-defined gesture set from all gestures proposed by all participants, we collated the gestures and asked participants to perform all the gestures, and conducted subjective ratings from 4 dimensions. We resolved repeatability between gestures by empirical categories, which intuitively characterized the similarity between gestures from 3 dimensions. We chose one gesture from each category to a subset of the most preferable gestures.

\subsubsection{Subjective Evaluation}
After deriving a gesture set with 15 gestures, we sought to find out which gesture is most suited for voice interactions, especially in social acceptance and using fatigue. We recruited 25 participants (10 male and 15 female) for our subjective evaluation, with an average age of 21(from 19 to 32, SD = 2.1). All of the participants were right-handed. Each participant performed all gestures three times using their right hand. The order of the gestures was pre-determined to counterbalance ordering effects. For each gesture, the experimenter would show an example video of this gesture to ensure the participant could perform the gesture correctly. The participant then followed the instructions provided on a laptop screen to perform gestures. After performing the gesture three times, the participant was asked to rate the gesture according to the following four criteria along a 7-point Likert scale (1: strongly disagree to 7: strongly agree), and the results are shown in Fig. \ref{fig:gestures}:
\begin{itemize}
    \item \textbf{Usability} measured ergonomics to reflect the comfort of the gesture. The participants are required to consider the gesture not only in stationary conditions (e.g., sitting) but also under moving conditions (e.g., running). The higher the score, the easier the gesture is to perform.
    \item \textbf{Social Acceptance} measures if the user will feel uncomfortable or embarrassed, or if performing the gesture will disturb others in public settings. The higher the score, the more acceptable the gesture is in social environments.
    \item \textbf{Disambiguity} measures the difficulty of confusing the gesture with daily hand movements or with other gestures. The higher the score, the less ambiguous the gesture is.
    \item \textbf{Fatigue} measures the physiological burden to perform the gesture. The higher the score, the less fatigue the gesture is to perform.
    
\end{itemize}

\subsubsection{Design Principles and Finalized Gesture Set}
In order to eliminate the design consistency and gestures with signal similarity to derive gestures that can be naturally performed and quickly remembered, we categorized all the gestures to propose the most representative one in each category. Considering the propagation path of the human voice around the head, we \revision{identify three structural properties to represent the proposed gesture set}, which are illustrated in Table \ref{tab:gestures}:
\begin{itemize}
    \item \textbf{Contact Position:} Due to the different contact positions of the fingers, the microphone mounted on the ring can receive different sounds. Because the mouth is the source of the sound, the closer the finger touches the sound source, the louder the sound will be picked up by the microphone on the ring. In all gestures, the contact positions of the fingers are the mouth(M), the cheek(CK), the chin(CN), and the ear(E).
    \item \textbf{Contact Type:} Different contact types, specifically divided into fingers(F), palms(P), and hand segments(HS) by which hands used to contact the face region, have clear distinctions in morphology which can be distinguished easily by users without ambiguity. Furthermore, different contact types will form a unique structure on the face to affect the collection of the earphones' feed-forward microphones. 
    \item \textbf{Occlusion State:} The occlusion state, which is separated in 3 levels(hardly(H), partially(P), and Completely(C), will produce different sounds by affecting the propagation path from the human voice to the ears. For example, gesture 7 (cover ear with arched palm) and gesture 12 (cover mouth with palm) shown in Fig. \ref{fig:gestures}, which 'completely' occlude the receiver (ears) and the transmitter (mouth) of the sound the propagation path in the air, will cause a loss of high-frequency sound.
\end{itemize}

% structural property
\revision{We combined the subjective evaluation results shown on Fig. \ref{fig:gestures} (in 4 dimensions: usability, social acceptance, disambiguity, and fatigue) with the gesture set optimization process. The process was based on first grouping gestures with the same structural property combinations from the above three dimensions and chose one gesture according to the subjective scores to represent each category.} From the categorization results, we found that gestures belonging to the E-F-H category include gestures 1, 9, 10, and 14. This type of gesture with fingers to contact the ear region and have similarity in signal. Moreover, E-F-H gestures are also commonly used to interact with earphones. We chose gesture 1 to represent E-F-H gestures in our final subset of VAHF gestures according to subjective ratings; Gestures 3,5 belong to M,CK-P-P gestures. \revision{Considering the operating region of gesture 4 is also near mouth and cheek(M, CK) although it does not actually contact the face and the gesture 4's other two dimensions are the same P(Contact Type)-P(Occulasion State) with gesture 3, 5, we grouped gestures 3,4,5 into one category and chose gesture 5 to represent this category.} Gestures 11 and 13 are omitted due to their lower social acceptance \revision{(e.g., lower than 3.5)}. \revision{And the remaining gestures (2,6,7,8,12,15) are kept in the gesture set due to their specificity in the three dimensions and higher subjective scores.} 
The gesture selection process resulted in 8 gestures. We checked the subjective scores of each dimension of the eight gestures selected again and found they are all above 4.4 and have a comprehensively higher score over others, which proves that our gesture selection is subjectively reasonable and practical for users. 

The above gesture selection process filtered out the following 8 gestures: gesture 1 (E-F-H), gesture 2 (M,CN-HS-H), gesture 5 (M,CK-P-P), gesture 6 (M-HS-C), gesture 7 (E-P-C), gesture 8 (M,CK-P-C), gesture 12 (M,CK,CN-P-C), gesture 15 (M,CK,CN,E-HS-P) where the indexes were consistent with Fig. \ref{fig:gestures}. These gestures constituted our final gesture set.

\section{Recognizing Voice-Accompanying Hand-to-Face Gestures with Cross-Device Sensing}

In this section, we introduce the design considerations and the technical details of our cross-device sensing method to recognize VAHF gestures. We explain the implementation considerations regarding device and channel selection. Then we present individual sensing models for vocal, ultrasonic, and IMU channels. Finally, we clarify the sensor combination and fusion strategies for real-world deployment.

\subsection{Considerations and Technical Overview}

We first clarify the considerations of our implementation before going into the technical details. To recognize VAHF gestures, we chose 3 types of commercial wearable devices - wireless ANC earbuds, smartwatches, and smart rings - as the sensor nodes in consideration of real-life deployment. Each wireless ANC earbud consists of an inner microphone and an outer microphone for noise canceling. The smartwatch is equipped with a microphone and a speaker which is capable to play sounds at 22.5 KHz and the ring is equipped with a microphone and an IMU. We chose the microphones and the IMUs as the sensor candidates in consideration of the computation efficiency for the always-availability (e.g., raise-to-speak technique on an Apple watch). These sensors are widely equipped on the aforementioned commercial wearable devices (earbuds, watch, and ring).

\begin{figure*}
    \centering
    \includegraphics[width=0.8\textwidth]{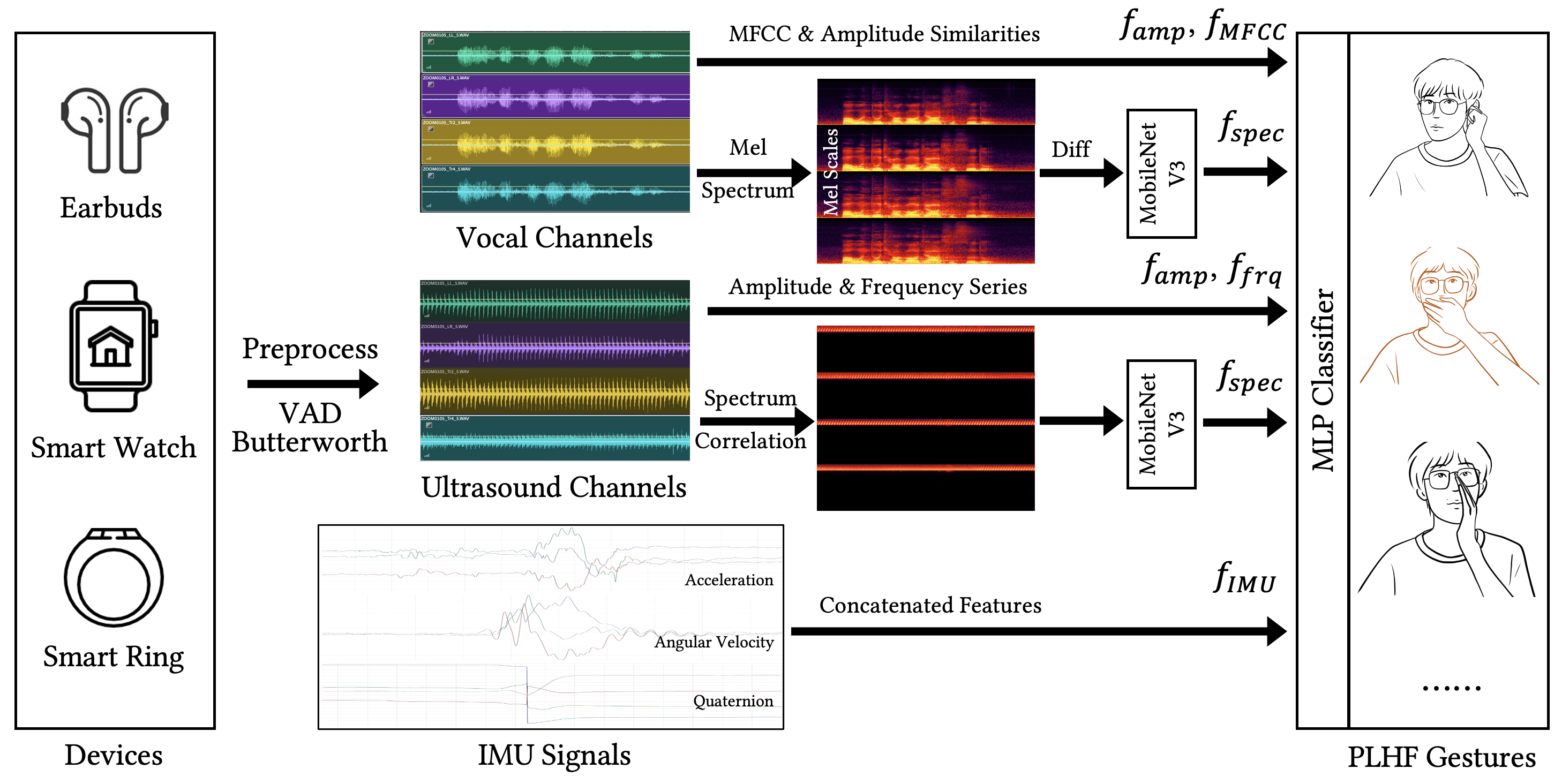}
    \caption{The sensing algorithm pipeline.}
    \Description{This figure contains the sensing system that consists of three independent models: vocal model, ultrasonic model, and IMU model. Each channel takes the corresponding preprocessed signal from the devices and outputs the feature vectors, which are fused and fed to the classifier layers to output the prediction logits.}
    \label{fig:algorithm}
\end{figure*}

An illustration of the entire system is shown in Figure \ref{fig:algorithm}. The sensing system consists of three independent models: vocal model, ultrasonic model, and IMU model. Each channel takes the corresponding preprocessed signal from the devices and outputs the feature vectors, which are fused and fed to the classifier layers to output the prediction logits.

% Each of the above devices is equipped with an inertial measurement unit (IMU) and at least one microphone. each earbud - 2 microphones.  The smart watch 
% is also equipped with a speaker which is capable to play sounds at 22.5 KHz(?). xxx, computational efficient, xx

% 我们的目标是识别语音交互中的hand-to-face手势，为了部署的可行性（能耗、性能、形态），我们选取了现有几种普及的商用可穿戴硬件设备 - 耳机、智能手表、智能戒指 - 作为可选的传感节点，并选取设备中两种常用的低功耗（computational feasibility）传感器 - 麦克风和IMU - 进行传感。

% 系统流水线如xx图，

\subsection{Recognizing VAHF Gestures with Single-Modality Solutions}

% Vocal, Ultrasonic, IMU

To facilitate efficient recognition of VAHF gestures, we first build three individual sensing models involving three independent channels of features - vocal features, ultrasonic features, and IMU features. Each channel of features serves as individual input of recognition in different dimensions.

\subsubsection{Vocal Model}

Performing hand-to-face gestures while speaking leads to changes in the acoustic property including amplitude, frequency response, and reverberation for the received signal. For example, an "hold up the palm beside nose and mouth" gesture may impede the direct transmission of sound to the left earbud's microphone, resulting in a lower amplitude and decay in high frequency in the corresponding channel. Since we focus on the difference in the acoustic property among the distributed microphones, we first figured out the reference channel. Typically, we chose the inner microphone as the reference channel when inner and outer microphones were simultaneously used and the outer microphone of the right earbud when inner channels were disabled.

%你这段是哪里的？我去看看 对就是内外耳 forward是外 back是内

Similar to prior work, we processes the audio data for classification using mel spectrum \cite{7952132}. Given a set of audio segments from all channels $[a_1,\cdots,a_n;a_{ref}]$ with the sample rate of 16 KHz, we convert each segment into the frequency domain by first applying the short-time Fourier transform then adopting a mel-scale transform with 128 mel filterbanks, after which we pad or trunk each spectrum in the temporal axis with zeros into $128 \times 250$ ($\approx 3$ seconds) and acquire $n+1$ maps $[m_1,\cdots,m_n;m_{ref}]$. Then we subtract the reference 
map $m_{ref}$ from all the monitored map $m_{i}$ to acquire the channel-wise difference in mel spectrum $[m_1 - m_{ref}, \cdots, m_n-m_{ref};m_{ref}]$. Finally, we concatenate all the maps in the first axis into a single input frame that can be fed into a deep-learning classification model. 

A MobileNet V3 Large \cite{Howard_2019_ICCV} model pretrained on ImageNet \cite{russakovsky2015imagenet} is used as the backbone network for feature extraction. The input frame is fed to the feature extractor layers of the pretrained model to generate a 1-D feature series $f_{spec}$. Such a model is chosen in consideration of the balance between computational complexity and performance \cite{Howard_2019_ICCV}. Benefitting from the well-designed network structure and the large parameter space with good initialization, MobileNet V3 Large has the potential in capturing fine-grained textural and geometry features from the concatenated spectrum map. 

Despite the direct use of the neural network on the mel frequency map, we extracted two additional sets of statistics features - transient signal amplitude and pair-wise similarity among Mel-frequency spectrum coefficients (MFCC) series - as classifier input, which is inspired by PrivateTalk's \cite{Yan-UIST-2019} solution in dealing with channel difference and delay between audio segments. For the transient amplitude feature, we use a sliding window with the size and stride of 200 to compute the amplitude series for each segment, after which we pad or trunk each series to a fixed length of 250. Then we concatenate all the amplitude series into a 1-D feature series $f_{amp}$. For pair-wise MFCC similarity, we first compute the MFCC series for each audio channel, then resample each MFCC map in the temporal domain into 20-frame segments with a stride of 10. For each pair of segment series, we compute their similarity using dynamic time warping (DTW) \cite{berndt1994using}. We acquired the pair-wise similarity feature vector $f_{MFCC}$ by concatenating all the above ${1\over2}n(n+1)$ simularity values.

After getting $f_{spec}$, $f_{amp}$, and $f_{MFCC}$, we concatenate them into a 1-dimensional vocal feature $f_{vol}$, which can either be used in an individual recognition model or be combined with other features. For an individual recognition model, $f_{vol}$ is fed into a multi-layer perceptron (MLP) classifier to predict the performed gesture.

% from the original audio segments
% statistics: signal amplitude, pair-wise distance DTW Inspired by PrivateTalk \cite{},

\subsubsection{Ultrasonic Model}

When the user performs a hand-to-face gesture, with his hand reaching different position on the face, the relative positions among the wrist, the finger, and the ears are temporally changing, thus yielding salient positional features. To facilitate such features, we devised an embedded ultrasonic sensing component, where the speaker on the smart watch works as an active source transmitting a 17.5 KHz - 22.5 KHz linear chirp signal which is captured by the microphones on the target devices. Such a design is inspired by the theory of Frequency Modulated Continuous Wave (FMCW) \cite{mao2016cat}, which is widely used in radar and indoor positioning systems to acquire positional tracking information. The sensing principals can be formalized as a typical linear chirp based FMCW. Let $x_0(t) = A_0 cos(2 \pi f_0 t + \pi {B \over T} t^2)$ be the source signal and $x_i(t) = A_1 cos(2 \pi f_0 (t-t_0) + \pi {B \over T} (t-t_0)^2)$ be the signal received by the $i^{th}$ device, where $B=f_1 - f_0$ is the bandwidth and $T$ is the period of the chirp. We first compute the correlation $x_0(t) x_i(t)$ and then pass the result to a low-pass filter to acquire the low-frequency component:

\begin{equation}
\begin{aligned}
    LPF(x_0(t)x_i(t)) = {1\over2} A_0 A_1 cos(2 \pi f_0 t_0 + \pi {B \over T} (2t_0 t - t_0^2))
\end{aligned}
\end{equation}

Note that $LPF(x_0(t)x_i(t))$ is a cosine function form of $t$ with an amplitude of ${1\over2} A_0 A_1$ and a frequency of $2{B \over T}t_0$, where the amplitude indicates the decay of the signal transmission and the frequency is proportional to the delay $t_0$ of the received signal. So we first compute the spectrum of $LPF(x_0(t)x_i(t))$ using short time Fourier transform (STFT). We extracted the following two features based on the spectrum: 1) the image feature $f_{spec}$ of the spectrum using a pretrained MobileNet V3 network and 2) the amplitude and frequency series $f_{stats}$ (which is flattened into a 1-D vector) derived from the spectrum. Finally, we concatenate $f_{spec}$ and $f_{stats}$ into $f_{ultra}$, which can be either fed into a downstream classifier or combined with other features as mentioned above.

\subsubsection{IMU Model}

Devices worn on the user's hand, such as a watch and ring, help to capture the movement and attitude of the user's moving hand, thus beneficial for recognizing hand-to-mouth gestures. In our setting, we choose to mount a 9-axis wireless IMU on the ring as previous work \cite{10.1145/3478114} did. The IMU reports 3-axis acceleration, 3-axis angular velocity, and the quaternion at 200 Hz. For gesture recognition, we adopted a fixed window of 400 frames (or 2 seconds), concatenating the acceleration, angular velocity, and quaternion series into a 4000-length vector. Then we used a 3-layer MLP with the structure of $Dropout(0.5) \rightarrow Linear(4000,512) \rightarrow ReLU \rightarrow Linear(512,512) \rightarrow ReLU \rightarrow Linear(512,9)$
for classification. 

\subsection{Sensor Combination and Fusion Strategies}

In consideration of the real-world deployment, we first figure out the reasonable device and sensor combinations. The devices include 1) left earbud (LE) with inner and outer microphones ($m_{l,i}$, $m_{l,o}$), 2) right earbud (RE) with inner and outer microphones ($m_{r,i}$, $m_{r,o}$), 3) watch with a microphone ($m_w$), and 4) ring with a microphone ($m_r$) and an IMU. Considering earbuds are most commonly used, we chose them as the primary device, which would work in different forms including two-side, one-side (wearing one earbud), and outer-only (for ones without active noise canceling). The introduction of the watch could be beneficial in providing an active ultrasound source as well as a hand-mounted microphone. Last, a ring device with an IMU and a microphone could track the movement of the hand and finger as well as provide a finger-mounted microphone. Based on the above observation, we devised four typical settings as follows for investigation: 1) single earbud (RE); 2) two earbuds (LE+RE); 3) two earbuds + watch (LE+RE+W); and 4) all devices (LE+RE+W+R).

%$m_{l,o}$, $m_{r,o}$, $m_w$, $m_r$

For settings 3) and 4), since the active ultrasound source and the IMU enable the ultrasonic model and the IMU model, a fusion method is required to fuse different recognition models from different channels. We investigated two fusion strategies: 1) logit-level fusion and 2) feature-level fusion. 

Let $F_{v}$, $F_{u}$, and $F_{i}$ be the feature extractor network of vocal, ultrasonic, and IMU models respectively and $C_{v}$, $C_{u}$, and $C_{i}$ be the corresponding multilayer classifier that outputs the logits. For logit-level fusion, the output logits are computed as 
\begin{equation}
    logits = a\cdot C_v(F_v(x_v)) + b\cdot C_u(F_u(x_u)) + c\cdot C_i(F_i(x_i))
\end{equation}
where $a$, $b$, and $c$ are learnable weight parameters ($x_v$, $x_u$, and $x_i$ are the corresponding channels of input). For feature-level fusion, the output logits are computed as 
\begin{equation}
    logits = C_{fuse}([F_v(x_v), F_u(x_u), F_i(x_i)])
\end{equation}
, where [*, *, *] refers to concatenation and $C_{fuse}$ is another MLP classifier that takes the concatenated features as input and outputs the logits.

% two fusion strategies:

% logit-level fusion

% feature-level fusion

% 4 different fusion strategies:

% 1) voting; 2) hierarchical; 3) feature-level fusion; 4) logit-level fusion; 5) + pretrained 

% \subsection{Towards Multi-Objective Optimization for xxxx}

% different targets:

% 1) optimizing recognition method
% 2) optimizing gesture set
% 3) optimizing form factor

%input/output
%不同的channels 
%device groups
%model

\section{Evaluation}
In this section, we conducted a systematic evaluation on the cross-device sensing method illustrated in the previous section. We first built a cross-device VAHF dataset consisting of \revision{10} users $\times$ 20 sentences $\times$ (8+1) gestures $=$ \revision{1800} samples. Then we evaluated our cross-device sensing method on the dataset in the dimensions of sensor combination, model selection, gesture reduction, and model ablation.

\subsection{Participants and Apparatus}
We recruited \revision{10} participants (\revision{6} female and 4 male) with an average age of  \revision{21.2 (from 19 to 28, SD=2.64)} and all participants were right-handed. All participants were recruited via emails and websites in our organization. 
We used a pair of earbuds with microphones, a smart watch with a motion sensor and a microphone, and a smart ring with a motion sensor and a microphone. The data of the microphones and motion sensors were fetched synchronously by a data collection thread. 

% 21 (from 19 to 28, SD=2.1)

\begin{figure*}
    \centering
    \includegraphics[width=0.8\textwidth]{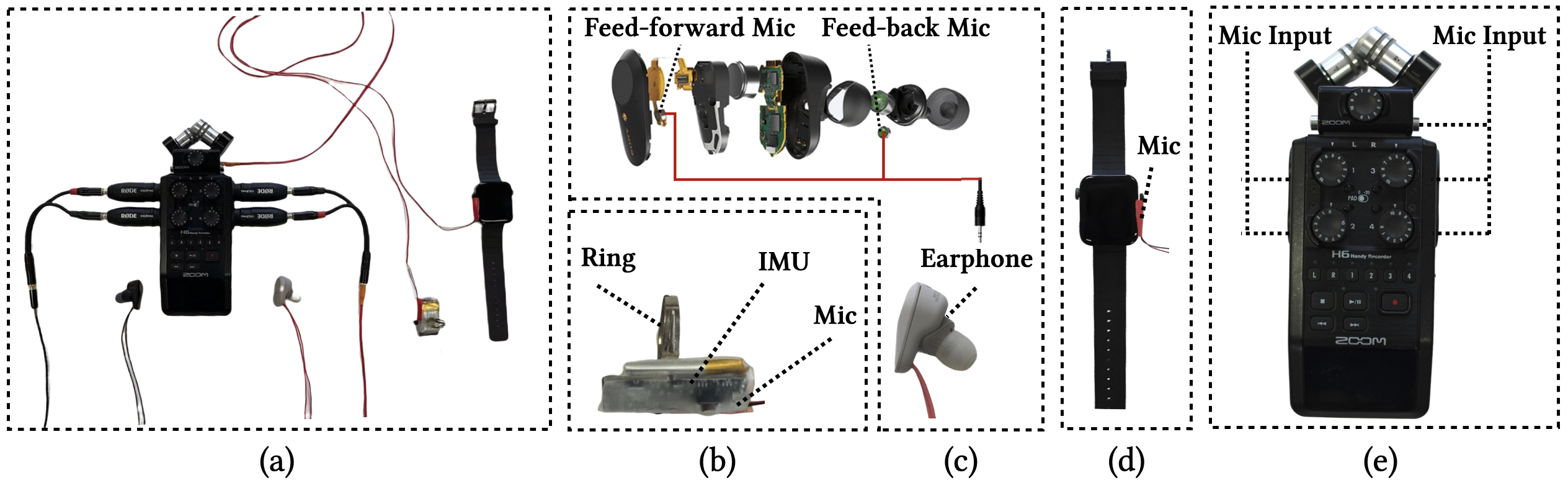}
    \caption{The apparatus of data collection. (a) Hardware overview. (b) A ring with an IMU and a microphone. (c) An earphone with a feed-forward microphone and a feed-back microphone. (d) A smartwatch with a microphone and a speaker. (e) A Zoom H6 recorder with six audio input channels.}
    \Description{This figure contains the apparatus of data collection. (a) Hardware overview. (b) A ring with an IMU and a microphone. (c) An earphone with a feed-forward microphone and a feed-back microphone. (d) A smartwatch with a microphone and a speaker. (e) A Zoom H6 recorder with six audio input channels.}
    \label{fig:hardware}
\end{figure*}

\begin{figure}
    \centering
    \includegraphics[width=1\columnwidth]{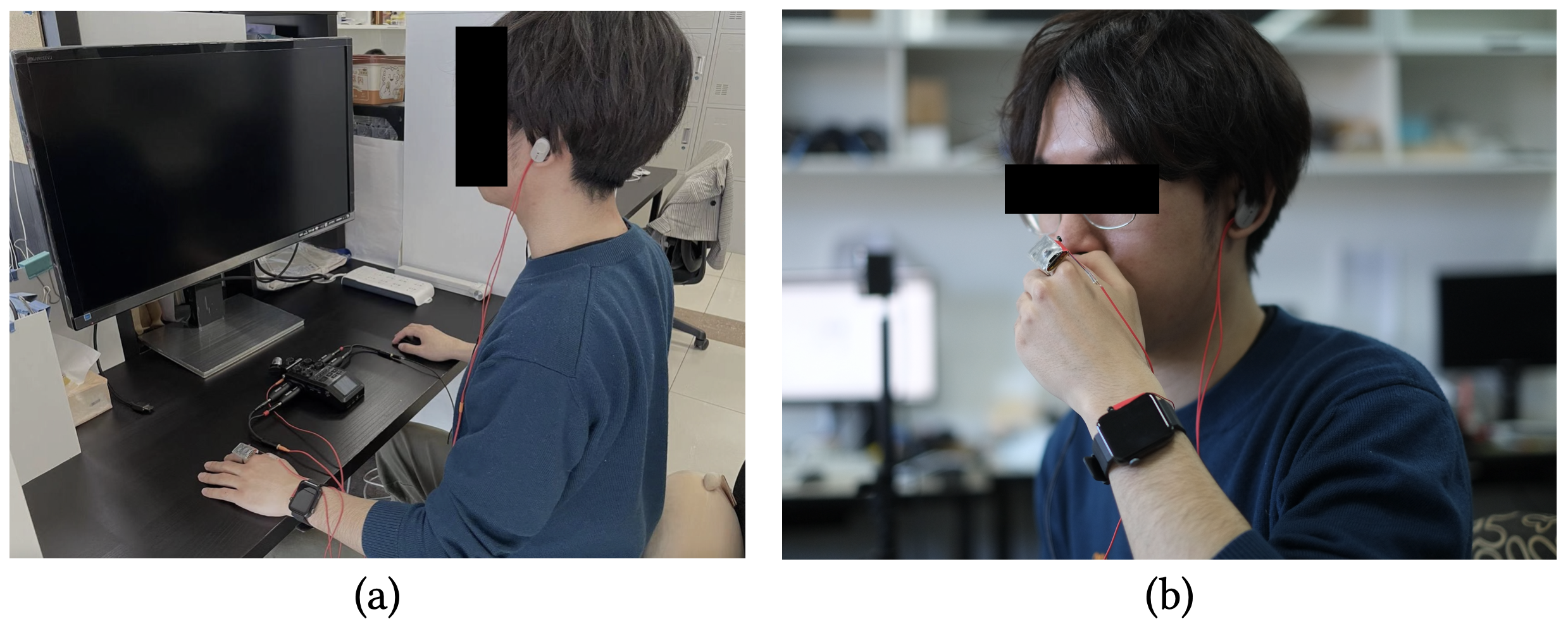}
    \caption{(a) Experiment setup. (b) A gesture example.}
    \Description{This figure contains the experiment setup and a gesture example. (a) The user was sitting at the table wearing all the devices. (b) The user performed a "cover mouth" gesture to record a data sample.}
    \label{fig:exp}
\end{figure}

\subsubsection{Microphones} We used the Sony WF-1000XM3 wireless noise-canceling headphone~\footnote{https://www.sony.com/electronics/truly-wireless/wf-1000xm3} and ZOOM H6 Handy recorder~\footnote{https://zoomcorp.com/en/us/handheld-recorders/handheld-recorders/h6-audio-recorder/} in this paper. We used four one-channel TRS audio cables to connect to the feed-forward and feed-back microphones
with headphones and a two-channel TRS audio cable to connect to the watch and the ring respectively. The ZOOM H6 audio recorder can record these six channels of timely synchronized audio data to a TF storage card (32 GB). The audio sampling rate was set to 48 KHz. To remain the same acoustic characteristic, we kept all the hardware in its original position in the earphone. The battery was run out of power to disable the on-chip software including the active noise canceling.

We use the MI Watch~\footnote{https://www.mi.com/global/mi-watch-lite} with 3-axis accelerometer and 3-axis gyroscope at 100Hz. The data is kept locally on watch and would be pulled after each round of the experiment. 

%超声波要不要写在这里
%这里也要插个图

\subsubsection{Inertial Measurement Unit}
We used a ring embedded with a wireless BMI-055 9-axis Inertial Measurement Unit (IMU) module, as shown in Figure \ref{fig:hardware}. The IMU data (3-axis acceleration, 3-axis gyroscope data, 3-axis geomagnetic data, and 3-axis Euler angle, current system time) is transmitted to a PC with a Bluetooth module at 200Hz (460800 baud rate).

\subsection{Data Collection}
We collected gesture samples from \revision{10} participants. The data collection entailed recording voice, ultrasound, and motion data while participants performed VAHF gestures corresponding to Section 3 and speak with daily voice commands. Each data collection study lasted about 60 minutes. Initially, participants were asked to read and sign consent forms. They were then shown instruction slides explaining the overall procedure of the data collection session and videos of VAHF gesture set from Section 3. Then we instructed participants to put on the earbuds, the smartwatch, and the ring properly, helping them to adjust the wearing until they felt comfortable with the devices. 
%文本不知道应该写在哪里,是不是要在sec3里面也写一下%. 

Each participant was required to perform 15 gestures and record 10 voice commands for each gesture (150 gesture samples in total). For each gesture, the participant was shown a slide with the gesture's name, the 10 voice commands, and the posture (sitting or standing). The order of the gestures and the posture condition were randomly picked to remove the order effect. The 10 voice commands were randomly picked from the daily Siri voice commands\footnote{https://support.apple.com/siri}, as shown in Table \ref{tab:commands}.

\begin{table}
  \vspace{-0.3cm}
  \centering
  \caption{The English translations of the voice commands used in data collection. The participants read the commands in Chinese. These voice commands were picked from Apple Siri's tutorial.}
  \Description{This table contains the English translations of the voice commands used in data collection. The participants read the commands in Chinese. These voice commands were picked from Apple Siri's tutorial. }
  ~\label{tab:commands}
    \vspace{-0.3cm}
    \resizebox{1\columnwidth}{!}{
    \begin{tabular}{l|l|l|l}
    \toprule
    Index & Voice Command & Index & Voice Command\\
    \midrule
    1  & Text Mom. & 11  & Turn the temperature up to 24 degrees.\\
    \midrule
    2  & Read my messages. & 12 & Show the photos taken today.\\
    \midrule
    3  & Who is calling? & 13 & Find the popular restaurants nearby. \\
    \midrule
    4 & Set an alarm for eight o'clock. & 14 & What is the latest movie?\\
    \midrule
    5 & Pay with Apple Pay. & 15 & How to take a holiday on National Day?\\
    \midrule
    6  & Transfer 20 yuan to Amy. & 16 & Buy train tickets to Beijing. \\
    \midrule
    7  & Remind me to pick up the clothes. & 17 & How is the weather today?  \\
    \midrule
    8  & What is my plan today? & 18 & Open Voice Memos. \\
    \midrule
    9  & Play my favorite song. & 19  & How to go to the nearest metro station? \\
    \midrule
    10  & Turn on the living room lights. & 20 & Countdown 20 minutes. \\
    \bottomrule
  \end{tabular}
  }
\end{table}

During the recording of the 10 gesture samples of each gesture, the experimenter first turned on the recording of the IMU ring, the watch's ultrasound, and the recorder. Then the participant clapped his or her hands to provide a synchronous signal used for the synchronization of different sensors. 
%没写第一个tick
For each gesture sample, the experimenter first pressed a key on the PC to label a tick and record the system time, which was used for gesture sample segmentation, and then signaled the participant to perform the gesture and read the corresponding voice command while keeping the gesture. This process was repeated 10 times until the participant finished all 10 gesture samples. After that, the experimenter turned off the recording.

%Once the experimenter pressed the start buttons for IMU ring, the watch's ultrasound, and the recorder, the participant would clap his or her hands to [provide a synchronous signal used for xxx] synchronize the different sensors[]. Then the experimenter would press the space on PC to indicate the participant to perform the desired gesture and record the current system time. [After the gesture prompt and execution,]x the participant [first performed and kept the desired gesture, and then read the corresponding voice command on the slide.] would keep the gesture and simultaneously read the corresponding voice command on the slide. The experimenter would press the space to label a tick on PC after the voice command and the participant would repeat the gesture and read the next voice command. The ticks were used for gesture sample segmentation. The time window between every space was labeled as a gesture segment. The labels were used to create the dataset and perform evaluations.

%用户在看到实验员按下按键之后，开始做手势，并在把手放在脸附近的时候开始读语音指令

\subsection{Data Preprocessing}
%synchronize
%segment
%label
%sample划分
%超声和人声的分离
Our data preprocessing process consisted of four steps: channel synchronization, segmentation, voice activation detection (VAD), and vocal-ultra sound separation. Below we illustrate the implementation details. 

\subsubsection{Channel Synchronization} The synchronization between audio channels was achieved by the audio card in ZOOM H6. To synchronize the audio channels and the IMU channel, we required the user to clap their hand to provide a signal for alignment before starting data collection for each session. Then we located such a clapping peak in both the audio channels and the IMU channel to acquire the relative time shift. The peak in the audio channels and the IMU channel was detected by finding the first local maximum in the amplitude spectrogram and the acceleration spectrogram, respectively.

\subsubsection{Audio Segmentation and VAD} After aligning all the channels, we segment the audio data which includes 10 voice commands for each. This was achieved by simply separating a piece of audio using the keystroke points annotated by the participants during recording. After getting the coarse segmentation, we further ran a VAD algorithm \cite{chang2006voice} to remove the silent period at the two ends in each segment.

%We need to segment audio data which includes 10 voice commands for each. After the gesture session starts, the user would clap his or her hands, and the applause would generate a peak in the amplitude spectrogram which is a second earlier than when the experimenter press the key to record the first system time point in average. So we took a second after the applause time as the starting point, and then use the remaining ten system time points to segment the audio. 

\subsubsection{Separation of Ultrasound and Vocals}
%butterworth highpass/lowpass
%1/2/3
We used a Butterworth~\footnote{https://en.wikipedia.org/wiki/Butterworth\textunderscore{}filter} highpass filter with 17500Hz cutoff frequency to separate the ultrasound and vocal from the audio data.

\subsection{Evaluation Design}

The evaluation consists of three sessions. In the first session, we conducted a two-factorial evaluation to analyze the recognition performance with regard to sensor combination and model selection. For sensor combination, corresponding to Section 4.3, we investigated five settings: 1) single (right) earbud with inner and outer microphones (RE, 2 audio channels), 2) two earbuds with inner and outer microphones (LE+RE, 4 audio channels), 3) two earbuds with outer microphones + watch (LE+RE+W, 3 audio channels), 4) all devices without the earbuds' inner channels (ALL-4ch, 4 audio channels), and 5) all devices with all channels (ALL-6ch, 6 audio channels). For model selection, we investigated the following six models: 1) vocal only (V), 2) ultrasound only (U), 3) IMU only (I), 4) vocal + ultrasound(V+U), 5) vocal + ultrasound + IMU with logit-level fusion (ALL-L), and 6) vocal + ultrasound + IMU with feature-level fusion (ALL-F). It is worth mentioning that the above two factors are correlated. The ultrasound channel would be activated unless the watch is used. Similarly, the IMU channel would be activated when the ring is used. Other factors, including the network structure, hyperparameters (max training epoch=100, dropout=0.5), and optimization strategies, are strictly controlled. We adopt three optimization strategies - pretraining, dropout, and \revision{warm-up} to improve the performance and training robustness of our model. For pretraining, we initialized the MobileNet V3's parameters with the one pretrained on ImageNet \cite{russakovsky2015imagenet}. For dropout, we added a dropout layer with a probability of 0.5 after the input layer to alleviate overfitting during training. For \revision{warm-up}, we adopted a \revision{warm-up} and weight decay strategy on the learning rate using the following piecewise function: if $n \leq 10$, then \revision{$lr(n)=0.1 \times n \times lr(0)$}, else $lr(n)=0.97^{n-10}\times lr(0)$, where $n$ is the training epoch and $lr(n)$ is the learning rate of the $n^{th}$ epoch.

In the second session, we conducted an extensive evaluation on a reduced gesture set to analyze the optimal performance and usability of each sensor combination for practical deployment. The reduced gesture set contains three signature gestures - cover mouth with palm (G1), cover ear with arched palm (G2), and hold up the palm beside nose and mouth (G3) - which received high preference scores from the previous user study and intuitively had significant effects on the acoustic propagation. For each sensor combination, we chose the optimal model, as acquired above, to compute the classification accuracies of each gesture and all three gestures (\{G1, E\}, \{G2, E\}, \{G3, E\}, and \{G1, G2, G3, E\}, where E refers to the empty gesture). All the evaluation settings were consistent with the first session. Such an evaluation helps to ground the applicability value of the minimal functionality under different hardware settings.

In the last session, we conducted an ablation study for the optimal model to analyze the effects of the optimization strategies in the model design: 1) pretraining, 2) dropout, and 3) warm-up. After getting the optimal model above, we ran the model in the same setting except disabling 1) all the three optimizations, 2) pretraining, 3) dropout, and 4) \revision{warm-up} to acquire the recognition accuracy in these 4 ablation settings. Such a study helped to validate the effectiveness of our model design.

All the above evaluations were conducted with leave-one-user-out cross-validation. \revision{For all the numerical comparisons, we reported the results along with the Wilcoxon Signed-Rank test to indicate the significance.}

% MODEL 4 PARTS: VOICE ULTRA BOOSTING IMU boosting
%优化目标：1.整体的正确率 2. 大类的可分的gestures 3.gesture的选取 3.有无boosting的 sensor-fusion/cross-device的组合 

\subsection{Results}

\begin{table}
  \vspace{-0.3cm}
  \centering
  \caption{Evaluation results regarding different sensor combinations and model selection. The notions in the table are consistent with Section 5.4. The numbers in the table indicate recognition accuracies in \% with standard deviations.}
  \Description{This table contains evaluation results regarding different sensor combinations and model selection. The notions in the table are consistent with Section 5.4. The numbers in the table indicate recognition accuracies in percent with standard deviations. }
  ~\label{tab:eval}
    \vspace{-0.3cm}
    \resizebox{1\columnwidth}{!}{
    \begin{tabular}{|l|p{40pt}|p{40pt}|p{40pt}|p{40pt}|p{40pt}|p{40pt}|}
    \toprule
     & V & U & I & V+U & ALL-L & ALL-F \\
    \midrule
    RE & 39.5(6.3) & - & - & - & - & -\\
    \midrule
    LE+RE & 70.3(10.0) & - & - & - & - & -\\
    \midrule
    LE+RE+W & 84.2(12.0) & 52.4(11.5) & - & 85.8(13.2) & - & -\\
    \midrule
    ALL-4ch & 89.9(10.5) & 66.7(13.6) & 49.0(5.3) & 89.8(10.3) & 90.8(10.2) & \textbf{91.5(8.9)}\\
    \midrule
    ALL-6ch & 90.0(10.9) & 70.9(14.5) & 49.0(5.3) & 89.2(11.4) & 90.7(9.9) & 90.9(9.4)\\
    \bottomrule
  \end{tabular}
  }
  \vspace{-0.2cm}
\end{table}

Table \ref{tab:eval} showed the results of the recognition performance regarding sensor combination and model selection. 

For vocal-only models, we observed a constant increase in recognition accuracy as more sensor nodes were introduced (e.g., from 39.5\% with a single earbud to 90.0\% with all the sensors, \revision{$Z=-2.81, p < 0.05$}). However, the difference between ALL-4ch and ALL-6ch was not significant, meaning when multiple devices were used, the introduction of the earbuds' inner channels brings limited information for the vocal channel. For ultrasonic-only models, the performance increased from 52.5\% to 70.9\% \revision{($Z=-2.81, p < 0.05$)} as the ring microphone and the earbuds' inner microphones were added. Notably, the independent use of the ultrasonic channels has its unique advantage of not relying on the vocal feature so that the model can still work well in scenarios such as noisy environments and whispering. The IMU model achieved an accuracy of 49.0\%, meaning the IMU could provide complementary information on hand and finger movement, though far from practical as an individual model. 

As for the sensor fusion models, we notice the vocal+ultra model had a performance increase over the vocal-only model with fewer input channels (LE+RE+W, \revision{84.2\% V.S. 85.8\%, $Z=-1.64, p=0.1$}), while it had no increase for ALL-4ch \revision{(89.9\% V.S. 89.8\%, $Z = -0.18, p = 0.86$)} and had a decrease for ALL-6ch \revision{(90.0\% V.S. 89.2\%, $Z = -0.98, p = 0.33$)}. This is probably because the vocal-only model with multiple channels (e.g., 6 channels) is a strong baseline, and combining it with an inferior model would introduce additional noise. Regarding all-channel fusion, we found feature-level fusion slightly outperformed logit-level fusion in accuracy \revision{(91.5\% V.S. 90.8\%, $Z = -0.98, p = 0.33$)}, probably due to the larger parameter space. We also observed a slight performance decrease for fusion models when adding the inner channels of the earbuds to ALL-4ch \revision{(91.5\% V.S. 90.9\%, $Z = -0.36, p = 0.72$)}, although the difference was not significant. The optimal model (all-channel feature-level fusion for ALL-4ch) achieved a 9-class recognition accuracy of 91.5\%\revision{, which significantly outperformed the vocal-only model ($Z=-1.96, p < 0.05$) with the same channels.}

% probably because of the redundancy of the inner channels

\begin{figure}
    \centering
    \includegraphics[width=1\columnwidth]{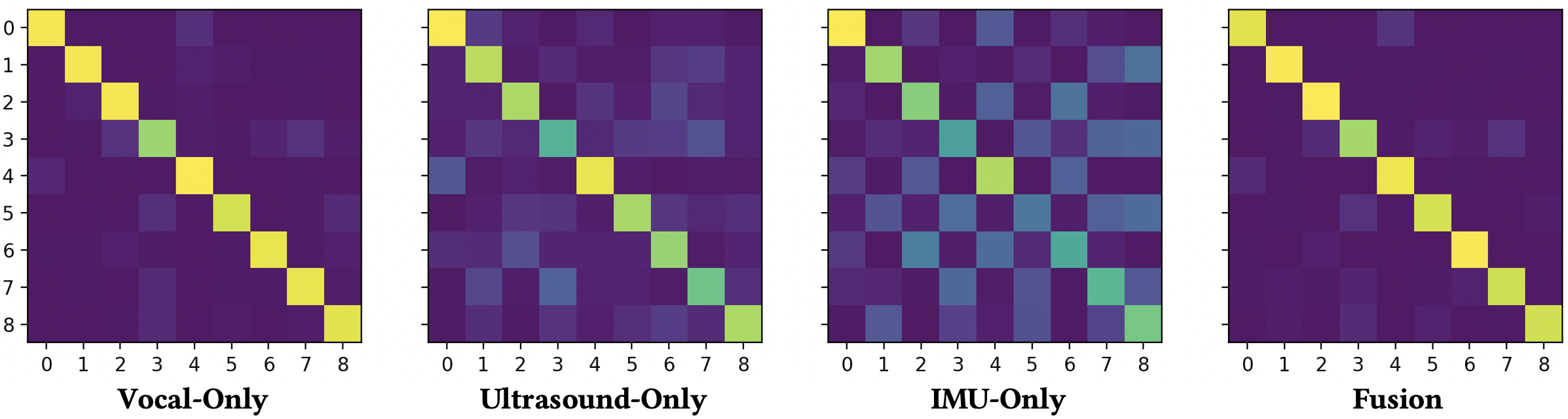}
    \caption{The confusion matrix of different models: vocal-only, ultra-only, IMU-only, and feature-level fusion. 0-9 represent the following gestures respectively: 0 - pinch the ear rim, 1 - calling gesture, 2 - support cheek with palm, 3 - cover mouth with palm, 4 - cover ear with arched palm, 5 - thinking face gesture, 6 - hold up the palm beside nose and mouth, 7 - cover mouth with fist, and 8 - empty gesture.}
    \Description{This figure contains the confusion matrix of different models: vocal-only, ultra-only, IMU-only, and feature-level fusion. 0-9 represent the following gestures respectively: 0 - pinch the ear rim, 1 - calling gesture, 2 - support cheek with palm, 3 - cover mouth with palm, 4 - cover ear with arched palm, 5 - thinking face gesture, 6 - hold up the palm beside nose and mouth, 7 - cover mouth with fist, and 8 - empty gesture. }
    \label{fig:confusion}
\end{figure}

To ground a better understanding of how each channel (vocal, ultrasound, and IMU) contributed to the recognition, we analyzed the confusion matrix of four models (vocal-only, ultra-only, IMU-only, and feature-level fusion) under ALL-4ch, as shown in Figure \ref{fig:confusion}. This result was understandable because for the gestures with larger confusion, we could easily figure out their similarity based on semantics. For example, gesture pairs $(0,4)$ and $(3,7)$ yield larger confusion for vocal and ultrasound models, where we observed similar touch positions for each pair of gestures (ear for $(0,4)$ and mouth for $(3,7)$). Gesture 1 confuses with gestures 3 and 7 in the ultrasound model probably due to a similar hand position, though it yields less confusion for the vocal model probably due to different occlusion levels (gestures 3 and 7 yield greater occlusion) that may influence the frequency response of the human voice. 

Results on the reduced gesture set were shown in Table \ref{tab:eval_reduced}. Since ALL-4ch achieved higher recognition accuracy than ALL-6ch in the fusion model (e.g., 91.5\% \revision{V.S.} 90.9\%), we dropped ALL-6ch in this table. We had the following observations: 1) Using one earbud with inner and outer microphones (RE), which is a severely restricted setting, could achieve a narrowly applicable accuracy of over $80\%$ for recognizing a specific single gesture (82.3\% for G1 and 83.0\% for G2) while it performed worse in recognizing other gesture (e.g., G3) or multiple gestures, which is understandable due to limited sensing information. 2) Using a pair of earbuds (LE+RE) could significantly boost the performance, with promising accuracies of 87.3\% for recognizing all three gestures and 98.8\% for recognizing G2, indicating the high applicability of such compact hardware form. 3) Additional hardware including a watch and ring brought the feasibility of fusing more input channels (e.g., ultrasound), which constantly improved the performance to a highly robust one (e.g., 97.3\% for recognizing all three gestures and 100\% for recognizing G2 and G3) and meanwhile lifting the distinguishable gesture space (e.g., from 3 gestures to 8 gestures, see Table \ref{tab:eval}) with high applicability (e.g., 91.5\% for simultaneously recognizing 8 gestures). The above results showed a leap over previous work with similar interaction modality (e.g., PrivateTalk \cite{Yan-UIST-2019}), revealing the feasibility of broadened gesture space (e.g., recognizing 8 gestures simultaneously) and the effectiveness of multi-device sensing.

\begin{table}
  \vspace{-0.3cm}
  \centering
  \caption{Recognition accuracy on the reduced gesture set. G1: cover mouth with palm, G2: cover ear with arched palm, and G3: hold up the palm beside nose and mouth.}
  \Description{This table contains the recognition accuracy on the reduced gesture set. G1: cover mouth with palm, G2: cover ear with arched palm, and G3: hold up the palm beside nose and mouth. }
  ~\label{tab:eval_reduced}
    \vspace{-0.3cm}
    \resizebox{1\columnwidth}{!}{
    \begin{tabular}{|l|p{40pt}|p{40pt}|p{40pt}|p{40pt}|}
    \toprule
     & RE & LE+RE & LE+RE+W & ALL-4ch \\
    \midrule
    G1 & 82.3(11.4) & 92.1(12.8) & 97.8(6.7) & 95.7(7.3) \\
    \midrule
    G2 & 83.0(13.0) & 98.8(1.9) & 97.9(6.3) & 100.0(0.0) \\
    \midrule
    G3 & 75.5(26.7) & 90.6(9.0) & 94.2(8.4) & 100.0(0.0) \\
    \midrule
    G1+G2+G3 & 64.4(10.6) & 87.3(8.9) & 91.3(11.3) & 97.3(4.5)\\
    \bottomrule
  \end{tabular}
  }
\end{table}

\begin{table}
  \centering
  \caption{Results of the ablation study. The numbers in the table indicate recognition accuracies in \% with standard deviations.}
  \Description{This table contains the results of the ablation study. The numbers in the table indicate recognition accuracies in \% with standard deviations. }
  ~\label{tab:ablation}
    \vspace{-0.3cm}
    
    \resizebox{0.5\columnwidth}{!}{
    \begin{tabular}{lc}
    \toprule
    Techniques & Accuracy \\
    \midrule
    No Optimization & 75.8(12.2) \\
    No Pretraining & 77.2(12.8) \\
    No Dropout & 91.2(7.8) \\
    No \revision{warm-up} & 87.8(11.1) \\
    \midrule
    Full Model & 91.5(8.9) \\
    \bottomrule
  \end{tabular}
  }
  \vspace{-0.3cm}
\end{table}

%  for single column
% \begin{table}
% \parbox{.62\linewidth}{
% \centering
% \caption{Recognition accuracy on the reduced gesture set. G1: cover mouth with palm, G2: cover ear with arched palm, and G3: hold up the palm beside nose and mouth.}~\label{tab:eval_reduced}
% \begin{tabular}{|l|p{40}|p{40}|p{40}|p{40}|}
%     \toprule
%      & RE & LE+RE & LE+RE+W & ALL-4ch \\
%     \midrule
%     G1 & 82.3(11.4) & 92.1(12.8) & 97.8(6.7) & 95.7(7.3) \\
%     \midrule
%     G2 & 83.0(13.0) & 98.8(1.9) & 97.9(6.3) & 100.0(0.0) \\
%     \midrule
%     G3 & 75.5(26.7) & 90.6(9.0) & 94.2(8.4) & 100.0(0.0) \\
%     \midrule
%     G1+G2+G3 & 64.4(10.6) & 87.3(8.9) & 91.3(11.3) & 97.3(4.5)\\
%     \bottomrule
%   \end{tabular}
% }
% \hfill
% \parbox{.33\linewidth}{
% \centering
% \caption{Results of the ablation study. The numbers in the table indicates recognition accuracies in \% with standard deviations.}~\label{tab:ablation}
% \begin{tabular}{lc}
%     \toprule
%     Techniques & Accuracy \\
%     \midrule
%     No Optimization & 75.8(12.2) \\
%     No Pretraining & 77.2(12.8) \\
%     No Dropout & 91.2(7.8) \\
%     No \revision{Warm-Up} & 87.8(11.1) \\
%     \midrule
%     Full Model & 91.5(8.9) \\
%     \bottomrule
%   \end{tabular}
% }
% \end{table}

The results of the ablation study are shown in Table \ref{tab:ablation}. We found disabling pretraining, dropout, and \revision{warm-up} caused different levels of performance degradation. Disabling pretraining caused the most significant decrease in performance ($-14.3\%$, \revision{$Z=-2.81, p < 0.05$}), which is probably because the feature extractor network (MobileNet V3) with pretraining on large-scale datasets could better extract different levels of image features. Meanwhile, disabling dropout caused slight decrease of $0.3\%$ \revision{($Z = -0.18, p = 0.86$), which was not significant,} and disabling \revision{warm-up} caused a decrease of $3.7\%$ \revision{($Z = -2.67, p < 0.05$)}. The introduction of \revision{warm-up} and dropout aims to optimize the training procedure (e.g., alleviating overfitting) and improve the robustness of the model. Compared with the raw model with no optimization, our model achieved a significant increase of $15.7\%$ \revision{($Z=-2.81, p < 0.05$)}, showing the superiority of all the optimization techniques. 
\section{Application Scenarios}
To demonstrate the applicability of VAHF gestures in voice interaction, we first presented the interaction space created by VAHF gestures along with example real-life scenarios. Then we discussed the design considerations and implications regarding the deployment of VAHF gestures in real practice.

% Then we conducted an informal study to understand users' preference and comments on XXX. Finally,

\subsection{Interaction Space and Scenario Description}
The introduction of VAHF gestures achieves the unique benefit of assigning a multi-class label to speech segments, which brings great potential to broaden the traditional voice interaction space in the following aspects.

\subsubsection{VAHF Gestures as Modality Control Signals.}

\textbf{Wakeup-free interaction.} The most intuitive function for modality control in voice interface is to use hand-to-face gestures (e.g., covering the mouth) to indicate whether the current speech is with interaction intention that should be processed by the voice assistant, which has been achieved and widely researched by previous work \cite{10.1145/3411764.3445687,Yan-UIST-2019,10.1145/3351276}. In our work, VAHF gestures have the inherited capability to support wakeup-free interaction simply by assigning one of the gestures for the wakeup state control.

\textbf{Dynamic modality control in multi-round interaction with voice assistant.} We demonstrate an example scenario using VAHF gestures for dynamic modality control in the multi-round dialog that has never been achieved before. When the user is enrolled in a multi-round dialog with the voice assistant, the complexity of the interaction behavior increases significantly. For example, in a specific dialog round, the user has different options to proceed with the dialog: 1) appending - the user appends a voice command and expects the voice assistant to process the command based on the dialog context in the regular order; 2) interrupting - the user wants to interrupt the current dialog (e.g., the voice assistant stops immediately and waits for new voice commands) and start a new dialog (abandoning the dialog context) with the new commands; and 3) editing - the user wants the voice assistant to edit the commands that they previously asked based on the dialog context and the brief editing command (e.g., the user says "How is the weather today?" When the assistant is answering, the user adds an editing command "No, I mean tomorrow."). Since our technique enables a channel width of up to 9 gestures (including the empty gesture) as modality input, we can assign different VAHF gestures to the three modalities of voice input - appending (e.g., covering the mouth), interrupting (e.g., covering the earphone), and editing (e.g., holding up the palm beside the mouth) - in the multi-round dialog scenario to enable more flexible and intelligent voice interaction flow. 

\subsubsection{Binding Shortcuts to VAHF Gestures}

\textbf{VAHF Gestures as UI shortcuts.} Simulating the execution of certain interaction paths through voice commands is a prevalent form of voice interaction on smartphones and wearable devices. When an interaction path takes a text entry slot or a period of raw speech as the input, it can be replaced with certain VAHF gestures. For example, the user can define the "phone call" VAHF gesture as opening WeChat and sending a voice message of the user's raw speech to Alice. Another example is to define the "thinking face" gesture as opening the Google website and searching for the text transcribed from the raw speech input. Such replacements of complex UI shortcuts with VAHF gestures could potentially reduce the repetition of the interaction path in speech, especially in a multi-round interaction.

\textbf{Registration and reservation of the VAHF-gesture-enabled shortcut session.} Regarding the binding of shortcuts with VAHF gestures, a more exciting design question is how the VAHF gestures are binded in real-world practice. Normally, the binding is fixed and can be set by the GUI (e.g., on a smartphone). On the contrary, we here present a dynamic registration and reservation mechanism for VAHF-gesture-enabled shortcut sessions, which are worthy of extensive exploration. In such a mechanism, for an unbinded VAHF gesture, when the user performs the gesture while narrating the full voice command, the voice assistant would automatically extract the UI shortcut from the command and bind it with the performed gesture. Later when the user wants to access the shortcut for a second time, they could perform the binded gesture while saying the input slot instead of the full command. The session and the dialog context are fully preserved for the gesture until a new command with UI shortcut semantics is input. The voice assistant would ask the user whether to update the binding of the gesture to a new shortcut. We believed such a design of a dynamic registration mechanism for VAHF gestures would benefit memorability, flexibility, and lower cold-start cost.

% that is worthy of extensive exploration The most interesting design is that the XXX can be dynamically registered and preserved 
%  (e.g., "opening Wechat and sending an emoji to Alice")

\subsubsection{VAHF Gestures as Spatial Indicators} VAHF gestures in voice interaction are also capable of indicating the target to interact with from the multiple interactable devices or elements. For example, in an IoT scenario where multiple voice-interactable devices (e.g., a smartphone, a TV, and a smart speaker) are in the same room, the user could perform different VAHF gestures with voice commands to trigger voice interaction with different devices. Similarly, in a complex UI control scenario (e.g., filling in a form with multiple text boxes), a VAHF gesture is displayed beside each text box, and the user could perform the corresponding VAHF gesture to input a particular text box.

\subsection{Design Considerations for VAHF Gestures to Enhance Voice Interaction}

The VAHF gestures proposed in our paper open the opportunity to design novel voice interactions for mobile, wearable, and HMD devices that allow users to quickly switch among modalities, accelerate common tasks, and manage multi-device interaction in different scenarios. We discussed two issues regarding the real-world deployment of VAHF gestures. \textbf{1) Combination strategy for better performance.} Although VAHF gestures have shown great potential in applicability, simply adding on all the functions described in the previous section is not feasible due to the channel capacity and the recognition accuracy. For example, as shown in Tables 3 and 4, an accuracy of 91.5\% for 9 classes is not yet highly usable, but a 4-class sub-gesture set achieved an accuracy of 97.3\%, which is considered highly usable. Therefore, a fine-grained design on the selection of gestures (e.g., alleviating using two gestures with higher confusion at the same time) and the switch of gesture sets in different scenarios is key to implementing a highly usable VAHF-gesture-enhanced voice interaction system. \textbf{2) Scalability and extensibility.} Although we only investigated an optimized VAHF gesture set with 8 gestures in our work, our sensing method was open to absorbing other extensive VAHF gestures. Our analysis method in Sections 3.1 and 3.2 provide a practical design guideline to elicit new gestures and analyze their feasibility. Further, our framework of recognizing VAHF gestures by multiple wearable devices has the advantage of appending or cutting down certain sensing channels easily, so the gesture set should be scalable and convertible for the system's flexibility.

% In this section, we first discuss two example real-life scenarios to demonstrate the applicability of PLHF gestures. Then we present a general design guideline for xxx.

% \subsection{Scenario Description}

% \subsubsection{}

% \subsubsection{}

% \subsection{Design Guideline} 
% increment of gesture or devices

% % 几种模式 
% 1. modality control, binding shortcuts
% 2. passing parameters, e.g., 
% 3. complex control, e.g., 

\section{Discussion and Limitations}
%noise effects
%ultrasound 
%ecological validity 

\subsection{Form Factor for Deployment}
Currently, our sensing algorithms were run \revision{and evaluated} in an offline setting with a full-functional prototype. \revision{As the instantiation, we also implemented a prototypical realtime VAHF gesture recognition system with the devices shown in Fig. \ref{fig:hardware}, a laptop, and a GPU server. The Zoom H6 recorder served as an audio card that streamed realtime audio data to the laptop. The PC ran a realtime data prepocessing program (written in Python, similar to Section 5.3) on one 2.3GHz Intel CPU core. The PC sent the processed audio segment to the GPU server using a sliding-window strategy while the recognition model processed the audio segment on one Nvidia RTX 3090 GPU and sent the results back to the PC. The whole pipeline ran at 60FPS with a delay of less than 50ms (excluding the network delay). The actual FPS could be controlled by adjusting the stride of the sliding window (typically an FPS higher than 5 could provide a good immediate experience). } 

Although our work demonstrated the computational feasibility of recognizing VAHF gestures, we should further consider the form factors for real-life deployment regarding synchronization, channel access, computational complexity, etc. We discussed the following three questions: 

\textbf{(1) How to synchronize and transmit the signal from different channels?} In our implementation, we used a strong synchronization system, where all the audio channels were wired to the audio card of ZOOM H6. In real deployment, the system could be implemented using Bluetooth low energy (BLE) technique for real-time signal transmission and synchronization (e.g., using broadcast mode or mesh mode for communication\footnote{https://www.bluetooth.com/learn-about-bluetooth/tech-overview/}), allowing dynamic communication among devices.

\textbf{(2) How to determine the proper channels or devices to enable in different scenarios?} As discussed in Section 6, we suggest the channels and devices should be enabled dynamically based on context information (e.g., the activated devices and the surrounding environment). The system would provide multiple levels of interaction progressively based on the activated devices (e.g., more complex gesture set for more devices) while preserving certain environment constraints (e.g., avoid using ultrasound in quiet scenarios or degrading the interaction capability in a noisy environment). With such context-aware optimizations, our technique could be implemented in a more user- and energy-friendly manner.

\textbf{(3) How to reduce the computational complexity?} \revision{In our implementation, we chose MobileNet V3, a light-weight NN model capable for mobile devices, as the backbone model in consideration of the computational efficiency. Further,} there are three possible ways to reduce the computational complexity: 1. using dynamic channels (e.g., using the minimal channels in an efficient mode); 2. using more light-weight feed-forward NN models (e.g., ShuffleNet\cite{zhang2018shufflenet}) for recognition; and 3. adopting bottom-level optimization (e.g., parameter quantization \cite{dettmers20158bit,courbariaux2014training} or customized hardware such as FPGA \cite{birem2014dreamcam}).

\subsection{Robustness against Environmental Interference}
Currently, our data were collected in an indoor environment with no background noise, aiming to validate the feasibility of recognizing VAHF gestures in an ideal setting. For real-world deployment, the recognition model is expected to deal with more complicated data with lower signal-noise ratio (SNR) and more environmental noise. So further research on the effect of environmental interference and how to build a robust recognition model should be conducted. Two strategies - 1) training the model with more diverse data coming from real-world scenarios or synthesization; 2) using advanced preprocessing techniques (e.g., active noise canceling algorithms) to reduce the noise and improve the SNR - may resolve this issue, which are worthy of further investigation. 

%Firstly, applications the proposed technique in the wild should be further conducted. Since our experiment was conducted in the lab with low environmental noise, we can expect that in a noisier environment (e.g., in a restaurant), the noise may overwhelm the voice input from users which may affect the performance of the algorithm. So a robust noise cancelling algorithm is needed.

\subsection{Ultrasound Usage}
We were well acknowledged the use of ultrasound for sensing could be controversial due to the interference and damage to one's hearing. In our work, we used a chirp signal from 17.5KHz to 22.5KHz and we noticed in our data collection procedure, some of the participants could hear the ultrasound and found it annoying. Further, ultrasound at sufficient sound pressure levels exert underlying danger of hearing damage even if it cannot be heard (though we strictly controlled the ultrasound amplitude in our study). Therefore, the use of ultrasound, including the amplitude, frequency, and duration should be more carefully designed for a gesture recognition system. More research should be conducted to explore the use of ultrasound and alternative sensing methods.

%noisy environment
%ecological validity of datasets
%

%design
%another sensors or devices? 
%ultrasound for human health?

\section{Conclusion}

In this paper, we investigated the design space and the recognition method of voice-accompanying hand-to-face (VAHF) gestures to enhance voice interaction with  parallel gesture channels. To design VAHF gestures, we first conducted an elicitation study, resulting in a total proposal of 15 gestures, followed by a hierarchical analysis process to output the most salient 8 gestures with the least ambiguity and physical confusion. Then we proposed a novel cross-device sensing method fusing different sensor channels to recognize para-linguistic hand-to-face gestures, achieving a high recognition accuracy of 97.3\% for 3+1\revision{(empty)} gestures and 91.5\% for 8+1\revision{(empty)} gestures recognition on our cross-device VAHF dataset. The uniqueness of our work is that we explored a broadened and scalable VAHF-gesture-based interaction space, which remains under-researched, to facilitate voice interaction in a more diverse manner (e.g., defining a shortcut or parsing parameters). Compared with prior work \cite{10.1145/3411764.3445687,Yan-UIST-2019} where a specific gesture (e.g., bringing the phone to the mouth\cite{10.1145/3411764.3445687}) was designed and recognized for 1-bit modality control (e.g., activating the voice assistant), our multi-device sensing framework is not only capable for recognizing up to 8 VAHF gestures simultaneously \revision{from the hand-off "empty" gesture}, but also benefits from the scalability (e.g., adding a device or adding a gesture is easy under our framework). As mobile devices and scenarios are becoming prevalent these years, voice input has become an essential modality of pervasive interaction. We envision our work would further enhance the efficiency and capability of current voice interaction and serve an important role in the future voice interaction of various scenarios like AR and IoT.

%% why not .tex

\begin{acks}
This work is supported by the Natural Science Foundation of China (NSFC) under Grant No. 62132010 and No. 62002198, Tsinghua University Initiative Scientific Research Program, Beijing Key Lab of Networked Multimedia, and Institute for Artificial Intelligence, Tsinghua University.
\end{acks}

\balance{}

\bibliographystyle{ACM-Reference-Format}
\bibliography{sample-base}


\begin{thebibliography}{86}

%%% the Web address.

\ifx \showCODEN    \undefined \def \showCODEN     #1{\unskip}     \fi
\ifx \showDOI      \undefined \def \showDOI       #1{#1}\fi
\ifx \showISBNx    \undefined \def \showISBNx     #1{\unskip}     \fi
\ifx \showISBNxiii \undefined \def \showISBNxiii  #1{\unskip}     \fi
\ifx \showISSN     \undefined \def \showISSN      #1{\unskip}     \fi
\ifx \showLCCN     \undefined \def \showLCCN      #1{\unskip}     \fi
\ifx \shownote     \undefined \def \shownote      #1{#1}          \fi
\ifx \showarticletitle \undefined \def \showarticletitle #1{#1}   \fi
\ifx \showURL      \undefined \def \showURL       {\relax}        \fi
% invisible to TeX
\providecommand\bibfield[2]{#2}
\providecommand\bibinfo[2]{#2}
\providecommand\natexlab[1]{#1}
\providecommand\showeprint[2][]{arXiv:#2}

\bibitem[\protect\citeauthoryear{Ahuja, Kong, Goel, and Harrison}{Ahuja
  et~al\mbox{.}}{2020}]%
        {10.1145/3379337.3415588}
\bibfield{author}{\bibinfo{person}{Karan Ahuja}, \bibinfo{person}{Andy Kong},
  \bibinfo{person}{Mayank Goel}, {and} \bibinfo{person}{Chris Harrison}.}
  \bibinfo{year}{2020}\natexlab{}.
\newblock \showarticletitle{Direction-of-Voice (DoV) Estimation for Intuitive
  Speech Interaction with Smart Devices Ecosystems}. In
  \bibinfo{booktitle}{\emph{Proceedings of the 33rd Annual ACM Symposium on
  User Interface Software and Technology}} (Virtual Event, USA)
  \emph{(\bibinfo{series}{UIST '20})}. \bibinfo{publisher}{Association for
  Computing Machinery}, \bibinfo{address}{New York, NY, USA},
  \bibinfo{pages}{1121–1131}.
\newblock
\showISBNx{9781450375146}
\urldef\tempurl%
\url{https://doi.org/10.1145/3379337.3415588}
\showDOI{\tempurl}


\bibitem[\protect\citeauthoryear{Ando, Kubo, Shizuki, and Takahashi}{Ando
  et~al\mbox{.}}{2017}]%
        {Ando2017}
\bibfield{author}{\bibinfo{person}{Toshiyuki Ando}, \bibinfo{person}{Yuki
  Kubo}, \bibinfo{person}{Buntarou Shizuki}, {and} \bibinfo{person}{Shin
  Takahashi}.} \bibinfo{year}{2017}\natexlab{}.
\newblock \showarticletitle{{CanalSense: Face-related movement recognition
  system based on sensing air pressure in ear canals}}.
\newblock \bibinfo{journal}{\emph{UIST 2017 - Proceedings of the 30th Annual
  ACM Symposium on User Interface Software and Technology}}
  (\bibinfo{year}{2017}), \bibinfo{pages}{679--689}.
\newblock
\showISBNx{9781450349819}
\urldef\tempurl%
\url{https://doi.org/10.1145/3126594.3126649}
\showDOI{\tempurl}


\bibitem[\protect\citeauthoryear{Berndt and Clifford}{Berndt and
  Clifford}{1994}]%
        {berndt1994using}
\bibfield{author}{\bibinfo{person}{Donald~J Berndt} {and}
  \bibinfo{person}{James Clifford}.} \bibinfo{year}{1994}\natexlab{}.
\newblock \showarticletitle{Using dynamic time warping to find patterns in time
  series.}. In \bibinfo{booktitle}{\emph{KDD workshop}},
  Vol.~\bibinfo{volume}{10}. Seattle, WA, USA:, \bibinfo{pages}{359--370}.
\newblock


\bibitem[\protect\citeauthoryear{Birem and Berry}{Birem and Berry}{2014}]%
        {birem2014dreamcam}
\bibfield{author}{\bibinfo{person}{Merwan Birem} {and}
  \bibinfo{person}{François Berry}.} \bibinfo{year}{2014}\natexlab{}.
\newblock \showarticletitle{DreamCam: A modular FPGA-based smart camera
  architecture}.
\newblock \bibinfo{journal}{\emph{Journal of Systems Architecture}}
  \bibinfo{volume}{60}, \bibinfo{number}{6} (\bibinfo{year}{2014}),
  \bibinfo{pages}{519--527}.
\newblock
\showISSN{1383-7621}
\urldef\tempurl%
\url{https://doi.org/10.1016/j.sysarc.2014.01.006}
\showDOI{\tempurl}


\bibitem[\protect\citeauthoryear{Bolt}{Bolt}{1980}]%
        {Bolt1980}
\bibfield{author}{\bibinfo{person}{Richard~A Bolt}.}
  \bibinfo{year}{1980}\natexlab{}.
\newblock \showarticletitle{{“Put-That-There”: Voice and Gesture at the
  Graphics Interface}}.
\newblock \bibinfo{journal}{\emph{SIGGRAPH Comput. Graph.}}
  \bibinfo{volume}{14}, \bibinfo{number}{3} (\bibinfo{year}{1980}),
  \bibinfo{pages}{262--270}.
\newblock
\showISSN{0097-8930}
\urldef\tempurl%
\url{https://doi.org/10.1145/965105.807503}
\showDOI{\tempurl}


\bibitem[\protect\citeauthoryear{Bourguet and Ando}{Bourguet and Ando}{1998}]%
        {Bourguet1998}
\bibfield{author}{\bibinfo{person}{Marie-Luce Bourguet} {and}
  \bibinfo{person}{Akio Ando}.} \bibinfo{year}{1998}\natexlab{}.
\newblock \showarticletitle{{Synchronization of Speech and Hand Gestures during
  Multimodal Human-Computer Interaction}}. In \bibinfo{booktitle}{\emph{CHI 98
  Conference Summary on Human Factors in Computing Systems}}
  \emph{(\bibinfo{series}{CHI '98})}. \bibinfo{publisher}{Association for
  Computing Machinery}, \bibinfo{address}{New York, NY, USA},
  \bibinfo{pages}{241--242}.
\newblock
\showISBNx{1581130287}
\urldef\tempurl%
\url{https://doi.org/10.1145/286498.286726}
\showDOI{\tempurl}


\bibitem[\protect\citeauthoryear{Buschek, Roppelt, and Alt}{Buschek
  et~al\mbox{.}}{2018}]%
        {7-arm}
\bibfield{author}{\bibinfo{person}{Daniel Buschek}, \bibinfo{person}{Bianka
  Roppelt}, {and} \bibinfo{person}{Florian Alt}.}
  \bibinfo{year}{2018}\natexlab{}.
\newblock \bibinfo{booktitle}{\emph{Extending Keyboard Shortcuts with Arm and
  Wrist Rotation Gestures}}.
\newblock \bibinfo{publisher}{Association for Computing Machinery},
  \bibinfo{address}{New York, NY, USA}, \bibinfo{pages}{1–12}.
\newblock
\showISBNx{9781450356206}
\urldef\tempurl%
\url{https://doi.org/10.1145/3173574.3173595}
\showURL{%
\tempurl}


\bibitem[\protect\citeauthoryear{Butler, Izadi, and Hodges}{Butler
  et~al\mbox{.}}{2008}]%
        {3-IR-touch-on-phone}
\bibfield{author}{\bibinfo{person}{Alex Butler}, \bibinfo{person}{Shahram
  Izadi}, {and} \bibinfo{person}{Steve Hodges}.}
  \bibinfo{year}{2008}\natexlab{}.
\newblock \showarticletitle{SideSight: Multi-"touch" Interaction around Small
  Devices}. In \bibinfo{booktitle}{\emph{Proceedings of the 21st Annual ACM
  Symposium on User Interface Software and Technology}} (Monterey, CA, USA)
  \emph{(\bibinfo{series}{UIST '08})}. \bibinfo{publisher}{Association for
  Computing Machinery}, \bibinfo{address}{New York, NY, USA},
  \bibinfo{pages}{201–204}.
\newblock
\showISBNx{9781595939753}
\urldef\tempurl%
\url{https://doi.org/10.1145/1449715.1449746}
\showDOI{\tempurl}


\bibitem[\protect\citeauthoryear{Ceolini, Frenkel, Shrestha, Taverni, Khacef,
  Payvand, and Donati}{Ceolini et~al\mbox{.}}{2020}]%
        {EMG-and-camera}
\bibfield{author}{\bibinfo{person}{Enea Ceolini}, \bibinfo{person}{Charlotte
  Frenkel}, \bibinfo{person}{Sumit~Bam Shrestha}, \bibinfo{person}{Gemma
  Taverni}, \bibinfo{person}{Lyes Khacef}, \bibinfo{person}{Melika Payvand},
  {and} \bibinfo{person}{Elisa Donati}.} \bibinfo{year}{2020}\natexlab{}.
\newblock \showarticletitle{Hand-gesture recognition based on EMG and
  event-based camera sensor fusion: A benchmark in neuromorphic computing}.
\newblock \bibinfo{journal}{\emph{Frontiers in Neuroscience}}
  \bibinfo{volume}{14} (\bibinfo{year}{2020}), \bibinfo{pages}{637}.
\newblock


\bibitem[\protect\citeauthoryear{Ceolini, Taverni, Khacef, Payvand, and
  Donati}{Ceolini et~al\mbox{.}}{2019}]%
        {EMG-and-camera2019}
\bibfield{author}{\bibinfo{person}{Enea Ceolini}, \bibinfo{person}{Gemma
  Taverni}, \bibinfo{person}{Lyes Khacef}, \bibinfo{person}{Melika Payvand},
  {and} \bibinfo{person}{Elisa Donati}.} \bibinfo{year}{2019}\natexlab{}.
\newblock \showarticletitle{Sensor fusion using EMG and vision for hand gesture
  classification in mobile applications}. In \bibinfo{booktitle}{\emph{2019
  IEEE Biomedical Circuits and Systems Conference (BioCAS)}}. IEEE,
  \bibinfo{pages}{1--4}.
\newblock


\bibitem[\protect\citeauthoryear{Chang, Kim, and Mitra}{Chang
  et~al\mbox{.}}{2006a}]%
        {chang2006voice}
\bibfield{author}{\bibinfo{person}{Joon-Hyuk Chang}, \bibinfo{person}{Nam~Soo
  Kim}, {and} \bibinfo{person}{Sanjit~K Mitra}.}
  \bibinfo{year}{2006}\natexlab{a}.
\newblock \showarticletitle{Voice activity detection based on multiple
  statistical models}.
\newblock \bibinfo{journal}{\emph{IEEE Transactions on Signal Processing}}
  \bibinfo{volume}{54}, \bibinfo{number}{6} (\bibinfo{year}{2006}),
  \bibinfo{pages}{1965--1976}.
\newblock


\bibitem[\protect\citeauthoryear{Chang, Kim, Lee, Cho, Soh, Shim, Yang, Cho,
  and Park}{Chang et~al\mbox{.}}{2006b}]%
        {1-capacitive}
\bibfield{author}{\bibinfo{person}{Wook Chang}, \bibinfo{person}{Kee~Eung Kim},
  \bibinfo{person}{Hyunjeong Lee}, \bibinfo{person}{Joon~Kee Cho},
  \bibinfo{person}{Byung~Seok Soh}, \bibinfo{person}{Jung~Hyun Shim},
  \bibinfo{person}{Gyunghye Yang}, \bibinfo{person}{Sung-Jung Cho}, {and}
  \bibinfo{person}{Joonah Park}.} \bibinfo{year}{2006}\natexlab{b}.
\newblock \showarticletitle{Recognition of grip-patterns by using capacitive
  touch sensors}. In \bibinfo{booktitle}{\emph{2006 IEEE International
  Symposium on Industrial Electronics}}, Vol.~\bibinfo{volume}{4}. IEEE,
  \bibinfo{pages}{2936--2941}.
\newblock


\bibitem[\protect\citeauthoryear{Chen, Xu, Li, Shi, Patel, and Wang}{Chen
  et~al\mbox{.}}{2021}]%
        {10.1145/3461778.3462004}
\bibfield{author}{\bibinfo{person}{Victor Chen}, \bibinfo{person}{Xuhai Xu},
  \bibinfo{person}{Richard Li}, \bibinfo{person}{Yuanchun Shi},
  \bibinfo{person}{Shwetak Patel}, {and} \bibinfo{person}{Yuntao Wang}.}
  \bibinfo{year}{2021}\natexlab{}.
\newblock \showarticletitle{Understanding the Design Space of Mouth
  Microgestures}. In \bibinfo{booktitle}{\emph{Designing Interactive Systems
  Conference 2021}} (Virtual Event, USA) \emph{(\bibinfo{series}{DIS '21})}.
  \bibinfo{publisher}{Association for Computing Machinery},
  \bibinfo{address}{New York, NY, USA}, \bibinfo{pages}{1068–1081}.
\newblock
\showISBNx{9781450384766}
\urldef\tempurl%
\url{https://doi.org/10.1145/3461778.3462004}
\showDOI{\tempurl}


\bibitem[\protect\citeauthoryear{Christofferson, Chen, Wang, Mariakakis, and
  Wang}{Christofferson et~al\mbox{.}}{2022}]%
        {Ken-sleep}
\bibfield{author}{\bibinfo{person}{Kenneth Christofferson},
  \bibinfo{person}{Xuyang Chen}, \bibinfo{person}{Zeyu Wang},
  \bibinfo{person}{Alex Mariakakis}, {and} \bibinfo{person}{Yuntao Wang}.}
  \bibinfo{year}{2022}\natexlab{}.
\newblock \showarticletitle{Sleep Sound Classification Using ANC-Enabled
  Earbuds}. In \bibinfo{booktitle}{\emph{2022 IEEE International Conference on
  Pervasive Computing and Communications Workshops and other Affiliated Events
  (PerCom Workshops)}}. \bibinfo{pages}{397--402}.
\newblock
\urldef\tempurl%
\url{https://doi.org/10.1109/PerComWorkshops53856.2022.9767394}
\showDOI{\tempurl}


\bibitem[\protect\citeauthoryear{Colli-Alfaro, Ibrahim, and
  Trejos}{Colli-Alfaro et~al\mbox{.}}{2019}]%
        {EMG-and-IMU-for-stroke}
\bibfield{author}{\bibinfo{person}{J.~Guillermo Colli-Alfaro},
  \bibinfo{person}{Anas Ibrahim}, {and} \bibinfo{person}{Ana~Luisa Trejos}.}
  \bibinfo{year}{2019}\natexlab{}.
\newblock \showarticletitle{Design of User-Independent Hand Gesture Recognition
  Using Multilayer Perceptron Networks and Sensor Fusion Techniques}. In
  \bibinfo{booktitle}{\emph{2019 IEEE 16th International Conference on
  Rehabilitation Robotics (ICORR)}}. \bibinfo{pages}{1103--1108}.
\newblock
\urldef\tempurl%
\url{https://doi.org/10.1109/ICORR.2019.8779533}
\showDOI{\tempurl}


\bibitem[\protect\citeauthoryear{Courbariaux, Bengio, and David}{Courbariaux
  et~al\mbox{.}}{2014}]%
        {courbariaux2014training}
\bibfield{author}{\bibinfo{person}{Matthieu Courbariaux},
  \bibinfo{person}{Yoshua Bengio}, {and} \bibinfo{person}{Jean-Pierre David}.}
  \bibinfo{year}{2014}\natexlab{}.
\newblock \bibinfo{title}{Training deep neural networks with low precision
  multiplications}.
\newblock
\newblock
\showeprint[arxiv]{1412.7024}~[cs.LG]


\bibitem[\protect\citeauthoryear{Dettmers}{Dettmers}{2015}]%
        {dettmers20158bit}
\bibfield{author}{\bibinfo{person}{Tim Dettmers}.}
  \bibinfo{year}{2015}\natexlab{}.
\newblock \bibinfo{title}{8-Bit Approximations for Parallelism in Deep
  Learning}.
\newblock
\newblock
\showeprint[arxiv]{1511.04561}~[cs.NE]


\bibitem[\protect\citeauthoryear{Fujie, Ejiri, Nakajima, Matsusaka, and
  Kobayashi}{Fujie et~al\mbox{.}}{2004}]%
        {1374748}
\bibfield{author}{\bibinfo{person}{S. Fujie}, \bibinfo{person}{Y. Ejiri},
  \bibinfo{person}{K. Nakajima}, \bibinfo{person}{Y. Matsusaka}, {and}
  \bibinfo{person}{T. Kobayashi}.} \bibinfo{year}{2004}\natexlab{}.
\newblock \showarticletitle{A conversation robot using head gesture recognition
  as para-linguistic information}. In \bibinfo{booktitle}{\emph{RO-MAN 2004.
  13th IEEE International Workshop on Robot and Human Interactive Communication
  (IEEE Catalog No.04TH8759)}}. \bibinfo{pages}{159--164}.
\newblock
\urldef\tempurl%
\url{https://doi.org/10.1109/ROMAN.2004.1374748}
\showDOI{\tempurl}


\bibitem[\protect\citeauthoryear{Fujie, Yamahata, and Kobayashi}{Fujie
  et~al\mbox{.}}{2006}]%
        {4115628}
\bibfield{author}{\bibinfo{person}{Shinya Fujie}, \bibinfo{person}{Toshihiko
  Yamahata}, {and} \bibinfo{person}{Tetsunori Kobayashi}.}
  \bibinfo{year}{2006}\natexlab{}.
\newblock \showarticletitle{Conversation Robot with the Function of Gaze
  Recognition}. In \bibinfo{booktitle}{\emph{2006 6th IEEE-RAS International
  Conference on Humanoid Robots}}. \bibinfo{pages}{364--369}.
\newblock
\urldef\tempurl%
\url{https://doi.org/10.1109/ICHR.2006.321298}
\showDOI{\tempurl}


\bibitem[\protect\citeauthoryear{Gao, Jin, Li, Choi, and Jin}{Gao
  et~al\mbox{.}}{2020}]%
        {Gao2020}
\bibfield{author}{\bibinfo{person}{Yang Gao}, \bibinfo{person}{Yincheng Jin},
  \bibinfo{person}{Jiyang Li}, \bibinfo{person}{Seokmin Choi}, {and}
  \bibinfo{person}{Zhanpeng Jin}.} \bibinfo{year}{2020}\natexlab{}.
\newblock \showarticletitle{{EchoWhisper: Exploring an Acoustic-Based Silent
  Speech Interface for Smartphone Users}}.
\newblock \bibinfo{journal}{\emph{Proc. ACM Interact. Mob. Wearable Ubiquitous
  Technol.}} \bibinfo{volume}{4}, \bibinfo{number}{3} (\bibinfo{year}{2020}).
\newblock
\urldef\tempurl%
\url{https://doi.org/10.1145/3411830}
\showDOI{\tempurl}


\bibitem[\protect\citeauthoryear{Gillian, Pfenninger, Russell, and
  Paradiso}{Gillian et~al\mbox{.}}{2014}]%
        {RFID}
\bibfield{author}{\bibinfo{person}{Nicholas Gillian}, \bibinfo{person}{Sara
  Pfenninger}, \bibinfo{person}{Spencer Russell}, {and}
  \bibinfo{person}{Joseph~A. Paradiso}.} \bibinfo{year}{2014}\natexlab{}.
\newblock \showarticletitle{Gestures Everywhere: A Multimodal Sensor Fusion and
  Analysis Framework for Pervasive Displays}. In
  \bibinfo{booktitle}{\emph{Proceedings of The International Symposium on
  Pervasive Displays}} (Copenhagen, Denmark) \emph{(\bibinfo{series}{PerDis
  '14})}. \bibinfo{publisher}{Association for Computing Machinery},
  \bibinfo{address}{New York, NY, USA}, \bibinfo{pages}{98–103}.
\newblock
\showISBNx{9781450329521}
\urldef\tempurl%
\url{https://doi.org/10.1145/2611009.2611032}
\showDOI{\tempurl}


\bibitem[\protect\citeauthoryear{Gong, Gupta, and Benko}{Gong
  et~al\mbox{.}}{2020}]%
        {10.1145/3379337.3415901}
\bibfield{author}{\bibinfo{person}{Jun Gong}, \bibinfo{person}{Aakar Gupta},
  {and} \bibinfo{person}{Hrvoje Benko}.} \bibinfo{year}{2020}\natexlab{}.
\newblock \bibinfo{booktitle}{\emph{Acustico: Surface Tap Detection and
  Localization Using Wrist-Based Acoustic TDOA Sensing}}.
\newblock \bibinfo{publisher}{Association for Computing Machinery},
  \bibinfo{address}{New York, NY, USA}, \bibinfo{pages}{406–419}.
\newblock
\showISBNx{9781450375146}
\urldef\tempurl%
\url{https://doi.org/10.1145/3379337.3415901}
\showURL{%
\tempurl}


\bibitem[\protect\citeauthoryear{Gu and Lien}{Gu and Lien}{2017}]%
        {7907299}
\bibfield{author}{\bibinfo{person}{Changzhan Gu} {and} \bibinfo{person}{Jaime
  Lien}.} \bibinfo{year}{2017}\natexlab{}.
\newblock \showarticletitle{A Two-Tone Radar Sensor for Concurrent Detection of
  Absolute Distance and Relative Movement for Gesture Sensing}.
\newblock \bibinfo{journal}{\emph{IEEE Sensors Letters}} \bibinfo{volume}{1},
  \bibinfo{number}{3} (\bibinfo{year}{2017}), \bibinfo{pages}{1--4}.
\newblock
\urldef\tempurl%
\url{https://doi.org/10.1109/LSENS.2017.2696520}
\showDOI{\tempurl}


\bibitem[\protect\citeauthoryear{Gu, Yu, Li, Li, Xu, Wei, and Shi}{Gu
  et~al\mbox{.}}{2019}]%
        {10.1145/3332165.3347947}
\bibfield{author}{\bibinfo{person}{Yizheng Gu}, \bibinfo{person}{Chun Yu},
  \bibinfo{person}{Zhipeng Li}, \bibinfo{person}{Weiqi Li},
  \bibinfo{person}{Shuchang Xu}, \bibinfo{person}{Xiaoying Wei}, {and}
  \bibinfo{person}{Yuanchun Shi}.} \bibinfo{year}{2019}\natexlab{}.
\newblock \showarticletitle{Accurate and Low-Latency Sensing of Touch Contact
  on Any Surface with Finger-Worn IMU Sensor}. In
  \bibinfo{booktitle}{\emph{Proceedings of the 32nd Annual ACM Symposium on
  User Interface Software and Technology}} (New Orleans, LA, USA)
  \emph{(\bibinfo{series}{UIST '19})}. \bibinfo{publisher}{Association for
  Computing Machinery}, \bibinfo{address}{New York, NY, USA},
  \bibinfo{pages}{1059–1070}.
\newblock
\showISBNx{9781450368162}
\urldef\tempurl%
\url{https://doi.org/10.1145/3332165.3347947}
\showDOI{\tempurl}


\bibitem[\protect\citeauthoryear{Gu, Yu, Li, Li, Wei, and Shi}{Gu
  et~al\mbox{.}}{2020}]%
        {Gu-IMU-RING-TYPING}
\bibfield{author}{\bibinfo{person}{Yizheng Gu}, \bibinfo{person}{Chun Yu},
  \bibinfo{person}{Zhipeng Li}, \bibinfo{person}{Zhaoheng Li},
  \bibinfo{person}{Xiaoying Wei}, {and} \bibinfo{person}{Yuanchun Shi}.}
  \bibinfo{year}{2020}\natexlab{}.
\newblock \showarticletitle{QwertyRing: Text Entry on Physical Surfaces Using a
  Ring}.
\newblock \bibinfo{journal}{\emph{Proc. ACM Interact. Mob. Wearable Ubiquitous
  Technol.}} \bibinfo{volume}{4}, \bibinfo{number}{4}, Article
  \bibinfo{articleno}{128} (\bibinfo{date}{Dec.} \bibinfo{year}{2020}),
  \bibinfo{numpages}{29}~pages.
\newblock
\urldef\tempurl%
\url{https://doi.org/10.1145/3432204}
\showDOI{\tempurl}


\bibitem[\protect\citeauthoryear{Gupta, Morris, Patel, and Tan}{Gupta
  et~al\mbox{.}}{2012}]%
        {gupta2012soundwave}
\bibfield{author}{\bibinfo{person}{Sidhant Gupta}, \bibinfo{person}{Dan
  Morris}, \bibinfo{person}{Shwetak~N. Patel}, {and} \bibinfo{person}{Desney
  Tan}.} \bibinfo{year}{2012}\natexlab{}.
\newblock \showarticletitle{{SoundWave: Using the Doppler effect to sense
  gestures}}. In \bibinfo{booktitle}{\emph{Conference on Human Factors in
  Computing Systems - Proceedings}}. \bibinfo{pages}{1911--1914}.
\newblock
\showISBNx{9781450310154}
\urldef\tempurl%
\url{https://doi.org/10.1145/2207676.2208331}
\showDOI{\tempurl}


\bibitem[\protect\citeauthoryear{Gustafson, Bierwirth, and Baudisch}{Gustafson
  et~al\mbox{.}}{2010}]%
        {17-interface}
\bibfield{author}{\bibinfo{person}{Sean Gustafson}, \bibinfo{person}{Daniel
  Bierwirth}, {and} \bibinfo{person}{Patrick Baudisch}.}
  \bibinfo{year}{2010}\natexlab{}.
\newblock \bibinfo{booktitle}{\emph{Imaginary Interfaces: Spatial Interaction
  with Empty Hands and without Visual Feedback}}.
\newblock \bibinfo{publisher}{Association for Computing Machinery},
  \bibinfo{address}{New York, NY, USA}, \bibinfo{pages}{3–12}.
\newblock
\showISBNx{9781450302715}
\urldef\tempurl%
\url{https://doi.org/10.1145/1866029.1866033}
\showURL{%
\tempurl}


\bibitem[\protect\citeauthoryear{Gustafson, Holz, and Baudisch}{Gustafson
  et~al\mbox{.}}{2011}]%
        {1-depth}
\bibfield{author}{\bibinfo{person}{Sean Gustafson}, \bibinfo{person}{Christian
  Holz}, {and} \bibinfo{person}{Patrick Baudisch}.}
  \bibinfo{year}{2011}\natexlab{}.
\newblock \showarticletitle{Imaginary Phone: Learning Imaginary Interfaces by
  Transferring Spatial Memory from a Familiar Device}. In
  \bibinfo{booktitle}{\emph{Proceedings of the 24th Annual ACM Symposium on
  User Interface Software and Technology}} (Santa Barbara, California, USA)
  \emph{(\bibinfo{series}{UIST '11})}. \bibinfo{publisher}{Association for
  Computing Machinery}, \bibinfo{address}{New York, NY, USA},
  \bibinfo{pages}{283–292}.
\newblock
\showISBNx{9781450307161}
\urldef\tempurl%
\url{https://doi.org/10.1145/2047196.2047233}
\showDOI{\tempurl}


\bibitem[\protect\citeauthoryear{Harrison, Schwarz, and Hudson}{Harrison
  et~al\mbox{.}}{2011}]%
        {Harrison2011}
\bibfield{author}{\bibinfo{person}{Chris Harrison}, \bibinfo{person}{Julia
  Schwarz}, {and} \bibinfo{person}{Scott~E Hudson}.}
  \bibinfo{year}{2011}\natexlab{}.
\newblock \showarticletitle{{TapSense: Enhancing Finger Interaction on Touch
  Surfaces}}. In \bibinfo{booktitle}{\emph{Proceedings of the 24th Annual ACM
  Symposium on User Interface Software and Technology}}
  \emph{(\bibinfo{series}{UIST '11})}. \bibinfo{publisher}{Association for
  Computing Machinery}, \bibinfo{address}{New York, NY, USA},
  \bibinfo{pages}{627--636}.
\newblock
\showISBNx{9781450307161}
\urldef\tempurl%
\url{https://doi.org/10.1145/2047196.2047279}
\showDOI{\tempurl}


\bibitem[\protect\citeauthoryear{Harrison, Xiao, and Hudson}{Harrison
  et~al\mbox{.}}{2012}]%
        {Harrison2012}
\bibfield{author}{\bibinfo{person}{Chris Harrison}, \bibinfo{person}{Robert
  Xiao}, {and} \bibinfo{person}{Scott~E. Hudson}.}
  \bibinfo{year}{2012}\natexlab{}.
\newblock \showarticletitle{{Acoustic barcodes: Passive, durable and
  inexpensive notched identification tags}}.
\newblock \bibinfo{journal}{\emph{UIST'12 - Proceedings of the 25th Annual ACM
  Symposium on User Interface Software and Technology}} (\bibinfo{year}{2012}),
  \bibinfo{pages}{563--567}.
\newblock
\showISBNx{9781450315807}


\bibitem[\protect\citeauthoryear{Hershey, Chaudhuri, Ellis, Gemmeke, Jansen,
  Moore, Plakal, Platt, Saurous, Seybold, Slaney, Weiss, and Wilson}{Hershey
  et~al\mbox{.}}{2017}]%
        {7952132}
\bibfield{author}{\bibinfo{person}{Shawn Hershey}, \bibinfo{person}{Sourish
  Chaudhuri}, \bibinfo{person}{Daniel P.~W. Ellis}, \bibinfo{person}{Jort~F.
  Gemmeke}, \bibinfo{person}{Aren Jansen}, \bibinfo{person}{R.~Channing Moore},
  \bibinfo{person}{Manoj Plakal}, \bibinfo{person}{Devin Platt},
  \bibinfo{person}{Rif~A. Saurous}, \bibinfo{person}{Bryan Seybold},
  \bibinfo{person}{Malcolm Slaney}, \bibinfo{person}{Ron~J. Weiss}, {and}
  \bibinfo{person}{Kevin Wilson}.} \bibinfo{year}{2017}\natexlab{}.
\newblock \showarticletitle{CNN architectures for large-scale audio
  classification}. In \bibinfo{booktitle}{\emph{2017 IEEE International
  Conference on Acoustics, Speech and Signal Processing (ICASSP)}}.
  \bibinfo{pages}{131--135}.
\newblock
\urldef\tempurl%
\url{https://doi.org/10.1109/ICASSP.2017.7952132}
\showDOI{\tempurl}


\bibitem[\protect\citeauthoryear{Hondori, Khademi, and Lopes}{Hondori
  et~al\mbox{.}}{2012}]%
        {Kinect-and-IMU}
\bibfield{author}{\bibinfo{person}{Hossein~Mousavi Hondori},
  \bibinfo{person}{Maryam Khademi}, {and} \bibinfo{person}{Cristina~V Lopes}.}
  \bibinfo{year}{2012}\natexlab{}.
\newblock \showarticletitle{Monitoring intake gestures using sensor fusion
  (microsoft kinect and inertial sensors) for smart home tele-rehab setting}.
  In \bibinfo{booktitle}{\emph{2012 1st Annual IEEE Healthcare Innovation
  Conference}}.
\newblock


\bibitem[\protect\citeauthoryear{Howard, Sandler, Chu, Chen, Chen, Tan, Wang,
  Zhu, Pang, Vasudevan, Le, and Adam}{Howard et~al\mbox{.}}{2019}]%
        {Howard_2019_ICCV}
\bibfield{author}{\bibinfo{person}{Andrew Howard}, \bibinfo{person}{Mark
  Sandler}, \bibinfo{person}{Grace Chu}, \bibinfo{person}{Liang-Chieh Chen},
  \bibinfo{person}{Bo Chen}, \bibinfo{person}{Mingxing Tan},
  \bibinfo{person}{Weijun Wang}, \bibinfo{person}{Yukun Zhu},
  \bibinfo{person}{Ruoming Pang}, \bibinfo{person}{Vijay Vasudevan},
  \bibinfo{person}{Quoc~V. Le}, {and} \bibinfo{person}{Hartwig Adam}.}
  \bibinfo{year}{2019}\natexlab{}.
\newblock \showarticletitle{Searching for MobileNetV3}. In
  \bibinfo{booktitle}{\emph{Proceedings of the IEEE/CVF International
  Conference on Computer Vision (ICCV)}}.
\newblock


\bibitem[\protect\citeauthoryear{Kashiwagi, Sugiura, Miyata, Tada, Sugimoto,
  and Saito}{Kashiwagi et~al\mbox{.}}{2017}]%
        {19-fish-eye}
\bibfield{author}{\bibinfo{person}{Naoaki Kashiwagi}, \bibinfo{person}{Yuta
  Sugiura}, \bibinfo{person}{Natsuki Miyata}, \bibinfo{person}{Mitsunori Tada},
  \bibinfo{person}{Maki Sugimoto}, {and} \bibinfo{person}{Hideo Saito}.}
  \bibinfo{year}{2017}\natexlab{}.
\newblock \showarticletitle{Measuring Grasp Posture Using an Embedded Camera}.
  In \bibinfo{booktitle}{\emph{2017 IEEE Winter Applications of Computer Vision
  Workshops (WACVW)}}. \bibinfo{pages}{42--47}.
\newblock
\urldef\tempurl%
\url{https://doi.org/10.1109/WACVW.2017.14}
\showDOI{\tempurl}


\bibitem[\protect\citeauthoryear{Katsamanis, Pitsikalis, Theodorakis, and
  Maragos}{Katsamanis et~al\mbox{.}}{2017}]%
        {Katsamanis2017}
\bibfield{author}{\bibinfo{person}{Athanasios Katsamanis},
  \bibinfo{person}{Vassilis Pitsikalis}, \bibinfo{person}{Stavros Theodorakis},
  {and} \bibinfo{person}{Petros Maragos}.} \bibinfo{year}{2017}\natexlab{}.
\newblock \bibinfo{booktitle}{\emph{{Multimodal Gesture Recognition}}}.
\newblock \bibinfo{publisher}{Association for Computing Machinery and Morgan
  {\&} Claypool}, \bibinfo{pages}{449--487}.
\newblock
\showISBNx{9781970001679}
\urldef\tempurl%
\url{https://doi.org/10.1145/3015783.3015796}
\showURL{%
\tempurl}


\bibitem[\protect\citeauthoryear{Kikuchi, Sugiura, Masai, Sugimoto, and
  Thomas}{Kikuchi et~al\mbox{.}}{2017}]%
        {EarTouch}
\bibfield{author}{\bibinfo{person}{Takashi Kikuchi}, \bibinfo{person}{Yuta
  Sugiura}, \bibinfo{person}{Katsutoshi Masai}, \bibinfo{person}{Maki
  Sugimoto}, {and} \bibinfo{person}{Bruce~H. Thomas}.}
  \bibinfo{year}{2017}\natexlab{}.
\newblock \showarticletitle{EarTouch: Turning the Ear into an Input Surface}.
  In \bibinfo{booktitle}{\emph{Proceedings of the 19th International Conference
  on Human-Computer Interaction with Mobile Devices and Services}} (Vienna,
  Austria) \emph{(\bibinfo{series}{MobileHCI '17})}.
  \bibinfo{publisher}{Association for Computing Machinery},
  \bibinfo{address}{New York, NY, USA}, Article \bibinfo{articleno}{27},
  \bibinfo{numpages}{6}~pages.
\newblock
\showISBNx{9781450350754}
\urldef\tempurl%
\url{https://doi.org/10.1145/3098279.3098538}
\showDOI{\tempurl}


\bibitem[\protect\citeauthoryear{Kim, Cha, Park, Nam, Lee, and Lee}{Kim
  et~al\mbox{.}}{2016}]%
        {Mo-Bi2016}
\bibfield{author}{\bibinfo{person}{Han-Jong Kim}, \bibinfo{person}{Seijin Cha},
  \bibinfo{person}{Richard~C. Park}, \bibinfo{person}{Tek-Jin Nam},
  \bibinfo{person}{Woohun Lee}, {and} \bibinfo{person}{Geehyuk Lee}.}
  \bibinfo{year}{2016}\natexlab{}.
\newblock \showarticletitle{Mo-Bi: Contextual Mobile Interfaces through
  Bimanual Posture Sensing with Wrist-Worn Devices}. In
  \bibinfo{booktitle}{\emph{Proceedings of HCI Korea}} (Jeongseon, Republic of
  Korea) \emph{(\bibinfo{series}{HCIK '16})}. \bibinfo{publisher}{Hanbit Media,
  Inc.}, \bibinfo{address}{Seoul, KOR}, \bibinfo{pages}{94–99}.
\newblock
\showISBNx{9788968487910}
\urldef\tempurl%
\url{https://doi.org/10.17210/hcik.2016.01.94}
\showDOI{\tempurl}


\bibitem[\protect\citeauthoryear{Laput, Ahuja, Goel, and Harrison}{Laput
  et~al\mbox{.}}{2018}]%
        {Laput2018}
\bibfield{author}{\bibinfo{person}{Gierad Laput}, \bibinfo{person}{Karan
  Ahuja}, \bibinfo{person}{Mayank Goel}, {and} \bibinfo{person}{Chris
  Harrison}.} \bibinfo{year}{2018}\natexlab{}.
\newblock \showarticletitle{{Ubicoustics: Plug-and-play acoustic activity
  recognition}}.
\newblock \bibinfo{journal}{\emph{UIST 2018 - Proceedings of the 31st Annual
  ACM Symposium on User Interface Software and Technology}}
  (\bibinfo{year}{2018}), \bibinfo{pages}{213--224}.
\newblock
\showISBNx{9781450359481}
\urldef\tempurl%
\url{https://doi.org/10.1145/3242587.3242609}
\showDOI{\tempurl}


\bibitem[\protect\citeauthoryear{Laput, Xiao, and Harrison}{Laput
  et~al\mbox{.}}{2016}]%
        {31-ViBand}
\bibfield{author}{\bibinfo{person}{Gierad Laput}, \bibinfo{person}{Robert
  Xiao}, {and} \bibinfo{person}{Chris Harrison}.}
  \bibinfo{year}{2016}\natexlab{}.
\newblock \showarticletitle{ViBand: High-Fidelity Bio-Acoustic Sensing Using
  Commodity Smartwatch Accelerometers}. In
  \bibinfo{booktitle}{\emph{Proceedings of the 29th Annual Symposium on User
  Interface Software and Technology}} (Tokyo, Japan)
  \emph{(\bibinfo{series}{UIST '16})}. \bibinfo{publisher}{Association for
  Computing Machinery}, \bibinfo{address}{New York, NY, USA},
  \bibinfo{pages}{321–333}.
\newblock
\showISBNx{9781450341899}
\urldef\tempurl%
\url{https://doi.org/10.1145/2984511.2984582}
\showDOI{\tempurl}


\bibitem[\protect\citeauthoryear{Lee, Lee, Shin, and Oakley}{Lee
  et~al\mbox{.}}{2018}]%
        {Acceptable}
\bibfield{author}{\bibinfo{person}{DoYoung Lee}, \bibinfo{person}{Youryang
  Lee}, \bibinfo{person}{Yonghwan Shin}, {and} \bibinfo{person}{Ian Oakley}.}
  \bibinfo{year}{2018}\natexlab{}.
\newblock \showarticletitle{Designing Socially Acceptable Hand-to-Face Input}.
  In \bibinfo{booktitle}{\emph{Proceedings of the 31st Annual ACM Symposium on
  User Interface Software and Technology}} (Berlin, Germany)
  \emph{(\bibinfo{series}{UIST '18})}. \bibinfo{publisher}{Association for
  Computing Machinery}, \bibinfo{address}{New York, NY, USA},
  \bibinfo{pages}{711–723}.
\newblock
\showISBNx{9781450359481}
\urldef\tempurl%
\url{https://doi.org/10.1145/3242587.3242642}
\showDOI{\tempurl}


\bibitem[\protect\citeauthoryear{Lee, Yeo, Dhuliawala, Akano, Shimizu, Starner,
  Quigley, Woo, and Kunze}{Lee et~al\mbox{.}}{2017}]%
        {nose}
\bibfield{author}{\bibinfo{person}{Juyoung Lee}, \bibinfo{person}{Hui-Shyong
  Yeo}, \bibinfo{person}{Murtaza Dhuliawala}, \bibinfo{person}{Jedidiah Akano},
  \bibinfo{person}{Junichi Shimizu}, \bibinfo{person}{Thad Starner},
  \bibinfo{person}{Aaron Quigley}, \bibinfo{person}{Woontack Woo}, {and}
  \bibinfo{person}{Kai Kunze}.} \bibinfo{year}{2017}\natexlab{}.
\newblock \showarticletitle{Itchy Nose: Discreet Gesture Interaction Using EOG
  Sensors in Smart Eyewear}. In \bibinfo{booktitle}{\emph{Proceedings of the
  2017 ACM International Symposium on Wearable Computers}} (Maui, Hawaii)
  \emph{(\bibinfo{series}{ISWC '17})}. \bibinfo{publisher}{Association for
  Computing Machinery}, \bibinfo{address}{New York, NY, USA},
  \bibinfo{pages}{94–97}.
\newblock
\showISBNx{9781450351881}
\urldef\tempurl%
\url{https://doi.org/10.1145/3123021.3123060}
\showDOI{\tempurl}


\bibitem[\protect\citeauthoryear{Li, Liang, Shrestha, Liu, Heidari, Le~Kernec,
  and Fioranelli}{Li et~al\mbox{.}}{2020}]%
        {UWB-and-Doppler}
\bibfield{author}{\bibinfo{person}{Haobo Li}, \bibinfo{person}{Xiangpeng
  Liang}, \bibinfo{person}{Aman Shrestha}, \bibinfo{person}{Yuchi Liu},
  \bibinfo{person}{Hadi Heidari}, \bibinfo{person}{Julien Le~Kernec}, {and}
  \bibinfo{person}{Francesco Fioranelli}.} \bibinfo{year}{2020}\natexlab{}.
\newblock \showarticletitle{Hierarchical Sensor Fusion for Micro-Gesture
  Recognition With Pressure Sensor Array and Radar}.
\newblock \bibinfo{journal}{\emph{IEEE Journal of Electromagnetics, RF and
  Microwaves in Medicine and Biology}} \bibinfo{volume}{4}, \bibinfo{number}{3}
  (\bibinfo{year}{2020}), \bibinfo{pages}{225--232}.
\newblock
\urldef\tempurl%
\url{https://doi.org/10.1109/JERM.2019.2949456}
\showDOI{\tempurl}


\bibitem[\protect\citeauthoryear{Liang, Hsia, Yu, Yan, Wang, and Shi}{Liang
  et~al\mbox{.}}{2022}]%
        {10.1145/3569463}
\bibfield{author}{\bibinfo{person}{Chen Liang}, \bibinfo{person}{Chi Hsia},
  \bibinfo{person}{Chun Yu}, \bibinfo{person}{Yukang Yan},
  \bibinfo{person}{Yuntao Wang}, {and} \bibinfo{person}{Yuanchun Shi}.}
  \bibinfo{year}{2022}\natexlab{}.
\newblock \showarticletitle{DRG-Keyboard: Enabling Subtle Gesture Typing on the
  Fingertip with Dual IMU Rings}.
\newblock \bibinfo{journal}{\emph{Proc. ACM Interact. Mob. Wearable Ubiquitous
  Technol.}} \bibinfo{volume}{6}, \bibinfo{number}{4}, Article
  \bibinfo{articleno}{170} (\bibinfo{date}{dec} \bibinfo{year}{2022}),
  \bibinfo{numpages}{30}~pages.
\newblock
\urldef\tempurl%
\url{https://doi.org/10.1145/3569463}
\showDOI{\tempurl}


\bibitem[\protect\citeauthoryear{Liang, Yu, Qin, Wang, and Shi}{Liang
  et~al\mbox{.}}{2021}]%
        {10.1145/3478114}
\bibfield{author}{\bibinfo{person}{Chen Liang}, \bibinfo{person}{Chun Yu},
  \bibinfo{person}{Yue Qin}, \bibinfo{person}{Yuntao Wang}, {and}
  \bibinfo{person}{Yuanchun Shi}.} \bibinfo{year}{2021}\natexlab{}.
\newblock \showarticletitle{DualRing: Enabling Subtle and Expressive Hand
  Interaction with Dual IMU Rings}.
\newblock \bibinfo{journal}{\emph{Proc. ACM Interact. Mob. Wearable Ubiquitous
  Technol.}} \bibinfo{volume}{5}, \bibinfo{number}{3}, Article
  \bibinfo{articleno}{115} (\bibinfo{date}{sep} \bibinfo{year}{2021}),
  \bibinfo{numpages}{27}~pages.
\newblock
\urldef\tempurl%
\url{https://doi.org/10.1145/3478114}
\showDOI{\tempurl}


\bibitem[\protect\citeauthoryear{Lien, Gillian, Karagozler, Amihood, Schwesig,
  Olson, Raja, and Poupyrev}{Lien et~al\mbox{.}}{2016}]%
        {10.1145/2897824.2925953}
\bibfield{author}{\bibinfo{person}{Jaime Lien}, \bibinfo{person}{Nicholas
  Gillian}, \bibinfo{person}{M.~Emre Karagozler}, \bibinfo{person}{Patrick
  Amihood}, \bibinfo{person}{Carsten Schwesig}, \bibinfo{person}{Erik Olson},
  \bibinfo{person}{Hakim Raja}, {and} \bibinfo{person}{Ivan Poupyrev}.}
  \bibinfo{year}{2016}\natexlab{}.
\newblock \showarticletitle{Soli: Ubiquitous Gesture Sensing with Millimeter
  Wave Radar}.
\newblock \bibinfo{journal}{\emph{ACM Trans. Graph.}} \bibinfo{volume}{35},
  \bibinfo{number}{4}, Article \bibinfo{articleno}{142} (\bibinfo{date}{July}
  \bibinfo{year}{2016}), \bibinfo{numpages}{19}~pages.
\newblock
\showISSN{0730-0301}
\urldef\tempurl%
\url{https://doi.org/10.1145/2897824.2925953}
\showDOI{\tempurl}


\bibitem[\protect\citeauthoryear{Liu, Nancel, and Vogel}{Liu
  et~al\mbox{.}}{2015}]%
        {2-depth}
\bibfield{author}{\bibinfo{person}{Mingyu Liu}, \bibinfo{person}{Mathieu
  Nancel}, {and} \bibinfo{person}{Daniel Vogel}.}
  \bibinfo{year}{2015}\natexlab{}.
\newblock \showarticletitle{Gunslinger: Subtle arms-down mid-air interaction}.
  In \bibinfo{booktitle}{\emph{Proceedings of the 28th Annual ACM Symposium on
  User Interface Software \& Technology}}. \bibinfo{pages}{63--71}.
\newblock


\bibitem[\protect\citeauthoryear{Liu, Shen, Li, and Wang}{Liu
  et~al\mbox{.}}{2019}]%
        {Liu2019}
\bibfield{author}{\bibinfo{person}{W Liu}, \bibinfo{person}{W Shen},
  \bibinfo{person}{B Li}, {and} \bibinfo{person}{L Wang}.}
  \bibinfo{year}{2019}\natexlab{}.
\newblock \showarticletitle{{Toward Device-Free Micro-Gesture Tracking via
  Accurate Acoustic Doppler-Shift Detection}}.
\newblock \bibinfo{journal}{\emph{IEEE Access}}  \bibinfo{volume}{7}
  (\bibinfo{year}{2019}), \bibinfo{pages}{1084--1094}.
\newblock
\showISSN{2169-3536}
\urldef\tempurl%
\url{https://doi.org/10.1109/ACCESS.2018.2886279}
\showDOI{\tempurl}


\bibitem[\protect\citeauthoryear{Liu, Zhang, and Gowda}{Liu
  et~al\mbox{.}}{2021}]%
        {2-EMG}
\bibfield{author}{\bibinfo{person}{Yilin Liu}, \bibinfo{person}{Shijia Zhang},
  {and} \bibinfo{person}{Mahanth Gowda}.} \bibinfo{year}{2021}\natexlab{}.
\newblock \showarticletitle{NeuroPose: 3D Hand Pose Tracking using EMG
  Wearables}. In \bibinfo{booktitle}{\emph{Proceedings of the Web Conference
  2021}}. \bibinfo{pages}{1471--1482}.
\newblock


\bibitem[\protect\citeauthoryear{Lu, Huang, Yu, Liu, and Shi}{Lu
  et~al\mbox{.}}{2020}]%
        {Lu-handtohand}
\bibfield{author}{\bibinfo{person}{Yiqin Lu}, \bibinfo{person}{Bingjian Huang},
  \bibinfo{person}{Chun Yu}, \bibinfo{person}{Guahong Liu}, {and}
  \bibinfo{person}{Yuanchun Shi}.} \bibinfo{year}{2020}\natexlab{}.
\newblock \showarticletitle{Designing and Evaluating Hand-to-Hand Gestures with
  Dual Commodity Wrist-Worn Devices}.
\newblock \bibinfo{journal}{\emph{Proc. ACM Interact. Mob. Wearable Ubiquitous
  Technol.}} \bibinfo{volume}{4}, \bibinfo{number}{1}, Article
  \bibinfo{articleno}{20} (\bibinfo{date}{March} \bibinfo{year}{2020}),
  \bibinfo{numpages}{27}~pages.
\newblock
\urldef\tempurl%
\url{https://doi.org/10.1145/3380984}
\showDOI{\tempurl}


\bibitem[\protect\citeauthoryear{Lv, Wang, and Shen}{Lv et~al\mbox{.}}{2011}]%
        {10.1145/2043674.2043723}
\bibfield{author}{\bibinfo{person}{Yanpeng Lv}, \bibinfo{person}{Shangfei
  Wang}, {and} \bibinfo{person}{Peijia Shen}.} \bibinfo{year}{2011}\natexlab{}.
\newblock \showarticletitle{A Real-Time Attitude Recognition by Eye-Tracking}.
  In \bibinfo{booktitle}{\emph{Proceedings of the Third International
  Conference on Internet Multimedia Computing and Service}} (Chengdu, China)
  \emph{(\bibinfo{series}{ICIMCS '11})}. \bibinfo{publisher}{Association for
  Computing Machinery}, \bibinfo{address}{New York, NY, USA},
  \bibinfo{pages}{170–173}.
\newblock
\showISBNx{9781450309189}
\urldef\tempurl%
\url{https://doi.org/10.1145/2043674.2043723}
\showDOI{\tempurl}


\bibitem[\protect\citeauthoryear{Mahmoud and Robinson}{Mahmoud and
  Robinson}{2011}]%
        {hand-over-gesture_book}
\bibfield{author}{\bibinfo{person}{Marwa Mahmoud} {and} \bibinfo{person}{Peter
  Robinson}.} \bibinfo{year}{2011}\natexlab{}.
\newblock \showarticletitle{Interpreting Hand-Over-Face Gestures}. In
  \bibinfo{booktitle}{\emph{Affective Computing and Intelligent Interaction}},
  \bibfield{editor}{\bibinfo{person}{Sidney D'Mello}, \bibinfo{person}{Arthur
  Graesser}, \bibinfo{person}{Bj{\"o}rn Schuller}, {and}
  \bibinfo{person}{Jean-Claude Martin}} (Eds.). \bibinfo{publisher}{Springer
  Berlin Heidelberg}, \bibinfo{address}{Berlin, Heidelberg},
  \bibinfo{pages}{248--255}.
\newblock
\showISBNx{978-3-642-24571-8}


\bibitem[\protect\citeauthoryear{Mao, He, and Qiu}{Mao et~al\mbox{.}}{2016}]%
        {mao2016cat}
\bibfield{author}{\bibinfo{person}{Wenguang Mao}, \bibinfo{person}{Jian He},
  {and} \bibinfo{person}{Lili Qiu}.} \bibinfo{year}{2016}\natexlab{}.
\newblock \showarticletitle{Cat: high-precision acoustic motion tracking}. In
  \bibinfo{booktitle}{\emph{Proceedings of the 22nd Annual International
  Conference on Mobile Computing and Networking}}. \bibinfo{pages}{69--81}.
\newblock


\bibitem[\protect\citeauthoryear{Mayer, Laput, and Harrison}{Mayer
  et~al\mbox{.}}{2020}]%
        {10.1145/3313831.3376479}
\bibfield{author}{\bibinfo{person}{Sven Mayer}, \bibinfo{person}{Gierad Laput},
  {and} \bibinfo{person}{Chris Harrison}.} \bibinfo{year}{2020}\natexlab{}.
\newblock \showarticletitle{Enhancing Mobile Voice Assistants with WorldGaze}.
  In \bibinfo{booktitle}{\emph{Proceedings of the 2020 CHI Conference on Human
  Factors in Computing Systems}} (Honolulu, HI, USA)
  \emph{(\bibinfo{series}{CHI '20})}. \bibinfo{publisher}{Association for
  Computing Machinery}, \bibinfo{address}{New York, NY, USA},
  \bibinfo{pages}{1–10}.
\newblock
\showISBNx{9781450367080}
\urldef\tempurl%
\url{https://doi.org/10.1145/3313831.3376479}
\showDOI{\tempurl}


\bibitem[\protect\citeauthoryear{McIntosh, McNeill, Fraser, Kerber,
  L{\"o}chtefeld, and Kr{\"u}ger}{McIntosh et~al\mbox{.}}{2016}]%
        {1-EMG}
\bibfield{author}{\bibinfo{person}{Jess McIntosh}, \bibinfo{person}{Charlie
  McNeill}, \bibinfo{person}{Mike Fraser}, \bibinfo{person}{Frederic Kerber},
  \bibinfo{person}{Markus L{\"o}chtefeld}, {and} \bibinfo{person}{Antonio
  Kr{\"u}ger}.} \bibinfo{year}{2016}\natexlab{}.
\newblock \showarticletitle{EMPress: practical hand gesture classification with
  wrist-mounted EMG and pressure sensing}. In
  \bibinfo{booktitle}{\emph{Proceedings of the 2016 CHI Conference on Human
  Factors in Computing Systems}}. \bibinfo{pages}{2332--2342}.
\newblock


\bibitem[\protect\citeauthoryear{Mesaros, Heittola, Eronen, and
  Virtanen}{Mesaros et~al\mbox{.}}{2010}]%
        {Mesaros2010}
\bibfield{author}{\bibinfo{person}{Annamaria Mesaros}, \bibinfo{person}{Toni
  Heittola}, \bibinfo{person}{Antti Eronen}, {and} \bibinfo{person}{Tuomas
  Virtanen}.} \bibinfo{year}{2010}\natexlab{}.
\newblock \showarticletitle{{Acoustic event detection in real life
  recordings}}.
\newblock \bibinfo{journal}{\emph{European Signal Processing Conference}}
  (\bibinfo{year}{2010}), \bibinfo{pages}{1267--1271}.
\newblock
\showISSN{22195491}


\bibitem[\protect\citeauthoryear{Moon, Yu, Wen, Shiratori, and Lee}{Moon
  et~al\mbox{.}}{2020}]%
        {moon2020interhand2}
\bibfield{author}{\bibinfo{person}{Gyeongsik Moon}, \bibinfo{person}{Shoou-I
  Yu}, \bibinfo{person}{He Wen}, \bibinfo{person}{Takaaki Shiratori}, {and}
  \bibinfo{person}{Kyoung~Mu Lee}.} \bibinfo{year}{2020}\natexlab{}.
\newblock \showarticletitle{Interhand2. 6m: A dataset and baseline for 3d
  interacting hand pose estimation from a single rgb image}. In
  \bibinfo{booktitle}{\emph{European Conference on Computer Vision}}. Springer,
  \bibinfo{pages}{548--564}.
\newblock


\bibitem[\protect\citeauthoryear{Morency, Sidner, Lee, and Darrell}{Morency
  et~al\mbox{.}}{2005}]%
        {10.1145/1088463.1088470}
\bibfield{author}{\bibinfo{person}{Louis-Philippe Morency},
  \bibinfo{person}{Candace Sidner}, \bibinfo{person}{Christopher Lee}, {and}
  \bibinfo{person}{Trevor Darrell}.} \bibinfo{year}{2005}\natexlab{}.
\newblock \showarticletitle{Contextual Recognition of Head Gestures}. In
  \bibinfo{booktitle}{\emph{Proceedings of the 7th International Conference on
  Multimodal Interfaces}} (Torento, Italy) \emph{(\bibinfo{series}{ICMI '05})}.
  \bibinfo{publisher}{Association for Computing Machinery},
  \bibinfo{address}{New York, NY, USA}, \bibinfo{pages}{18–24}.
\newblock
\showISBNx{1595930280}
\urldef\tempurl%
\url{https://doi.org/10.1145/1088463.1088470}
\showDOI{\tempurl}


\bibitem[\protect\citeauthoryear{Mueller, Bernard, Sotnychenko, Mehta, Sridhar,
  Casas, and Theobalt}{Mueller et~al\mbox{.}}{2018}]%
        {30-Generate-hands-tracking}
\bibfield{author}{\bibinfo{person}{Franziska Mueller}, \bibinfo{person}{Florian
  Bernard}, \bibinfo{person}{Oleksandr Sotnychenko}, \bibinfo{person}{Dushyant
  Mehta}, \bibinfo{person}{Srinath Sridhar}, \bibinfo{person}{Dan Casas}, {and}
  \bibinfo{person}{Christian Theobalt}.} \bibinfo{year}{2018}\natexlab{}.
\newblock \showarticletitle{GANerated Hands for Real-Time 3D Hand Tracking From
  Monocular RGB}. In \bibinfo{booktitle}{\emph{Proceedings of the IEEE
  Conference on Computer Vision and Pattern Recognition (CVPR)}}.
\newblock


\bibitem[\protect\citeauthoryear{Ono, Shizuki, and Tanaka}{Ono
  et~al\mbox{.}}{2013}]%
        {Ono2013}
\bibfield{author}{\bibinfo{person}{Makoto Ono}, \bibinfo{person}{Buntarou
  Shizuki}, {and} \bibinfo{person}{Jiro Tanaka}.}
  \bibinfo{year}{2013}\natexlab{}.
\newblock \showarticletitle{{Touch {\&} activate: adding interactivity to
  existing objects using active acoustic sensing}}.
\newblock \bibinfo{journal}{\emph{Proceedings of the 26th annual ACM symposium
  on User interface software and technology}} (\bibinfo{year}{2013}),
  \bibinfo{pages}{31--40}.
\newblock
\showISBNx{9781450322683}
\urldef\tempurl%
\url{https://doi.org/10.1145/2501988.2501989}
\showDOI{\tempurl}


\bibitem[\protect\citeauthoryear{Qin, Yu, Li, Zhong, Yan, and Shi}{Qin
  et~al\mbox{.}}{2021}]%
        {10.1145/3411764.3445687}
\bibfield{author}{\bibinfo{person}{Yue Qin}, \bibinfo{person}{Chun Yu},
  \bibinfo{person}{Zhaoheng Li}, \bibinfo{person}{Mingyuan Zhong},
  \bibinfo{person}{Yukang Yan}, {and} \bibinfo{person}{Yuanchun Shi}.}
  \bibinfo{year}{2021}\natexlab{}.
\newblock \showarticletitle{ProxiMic: Convenient Voice Activation via
  Close-to-Mic Speech Detected by a Single Microphone}. In
  \bibinfo{booktitle}{\emph{Proceedings of the 2021 CHI Conference on Human
  Factors in Computing Systems}} (Yokohama, Japan) \emph{(\bibinfo{series}{CHI
  '21})}. \bibinfo{publisher}{Association for Computing Machinery},
  \bibinfo{address}{New York, NY, USA}, Article \bibinfo{articleno}{8},
  \bibinfo{numpages}{12}~pages.
\newblock
\showISBNx{9781450380966}
\urldef\tempurl%
\url{https://doi.org/10.1145/3411764.3445687}
\showDOI{\tempurl}


\bibitem[\protect\citeauthoryear{Russakovsky, Deng, Su, Krause, Satheesh, Ma,
  Huang, Karpathy, Khosla, Bernstein, et~al\mbox{.}}{Russakovsky
  et~al\mbox{.}}{2015}]%
        {russakovsky2015imagenet}
\bibfield{author}{\bibinfo{person}{Olga Russakovsky}, \bibinfo{person}{Jia
  Deng}, \bibinfo{person}{Hao Su}, \bibinfo{person}{Jonathan Krause},
  \bibinfo{person}{Sanjeev Satheesh}, \bibinfo{person}{Sean Ma},
  \bibinfo{person}{Zhiheng Huang}, \bibinfo{person}{Andrej Karpathy},
  \bibinfo{person}{Aditya Khosla}, \bibinfo{person}{Michael Bernstein},
  {et~al\mbox{.}}} \bibinfo{year}{2015}\natexlab{}.
\newblock \showarticletitle{Imagenet large scale visual recognition challenge}.
\newblock \bibinfo{journal}{\emph{International journal of computer vision}}
  \bibinfo{volume}{115}, \bibinfo{number}{3} (\bibinfo{year}{2015}),
  \bibinfo{pages}{211--252}.
\newblock


\bibitem[\protect\citeauthoryear{Sapi{\'{n}}ski, Kami{\'{n}}ska, Pelikant,
  Ozcinar, Avots, and Anbarjafari}{Sapi{\'{n}}ski et~al\mbox{.}}{2019}]%
        {10.1007/978-3-030-05792-3_15}
\bibfield{author}{\bibinfo{person}{Tomasz Sapi{\'{n}}ski},
  \bibinfo{person}{Dorota Kami{\'{n}}ska}, \bibinfo{person}{Adam Pelikant},
  \bibinfo{person}{Cagri Ozcinar}, \bibinfo{person}{Egils Avots}, {and}
  \bibinfo{person}{Gholamreza Anbarjafari}.} \bibinfo{year}{2019}\natexlab{}.
\newblock \showarticletitle{Multimodal Database of Emotional Speech, Video and
  Gestures}. In \bibinfo{booktitle}{\emph{Pattern Recognition and Information
  Forensics}}, \bibfield{editor}{\bibinfo{person}{Zhaoxiang Zhang},
  \bibinfo{person}{David Suter}, \bibinfo{person}{Yingli Tian},
  \bibinfo{person}{Alexandra Branzan~Albu}, \bibinfo{person}{Nicolas
  Sid{\`e}re}, {and} \bibinfo{person}{Hugo Jair~Escalante}} (Eds.).
  \bibinfo{publisher}{Springer International Publishing},
  \bibinfo{address}{Cham}, \bibinfo{pages}{153--163}.
\newblock
\showISBNx{978-3-030-05792-3}


\bibitem[\protect\citeauthoryear{Sauras-Perez, Gil, {Singh Gill}, Pisu, and
  Taiber}{Sauras-Perez et~al\mbox{.}}{2017}]%
        {VoGe2017}
\bibfield{author}{\bibinfo{person}{Pablo Sauras-Perez}, \bibinfo{person}{Andrea
  Gil}, \bibinfo{person}{Jasprit {Singh Gill}}, \bibinfo{person}{Pierluigi
  Pisu}, {and} \bibinfo{person}{Joachim Taiber}.}
  \bibinfo{year}{2017}\natexlab{}.
\newblock \showarticletitle{{VoGe: A Voice and Gesture System for Interacting
  with Autonomous Cars}}. In \bibinfo{booktitle}{\emph{WCX™ 17: SAE World
  Congress Experience}}. \bibinfo{publisher}{SAE International}.
\newblock
\showISSN{0148-7191}
\urldef\tempurl%
\url{https://doi.org/10.4271/2017-01-0068}
\showDOI{\tempurl}


\bibitem[\protect\citeauthoryear{Serrano, Ens, and Irani}{Serrano
  et~al\mbox{.}}{2014}]%
        {HWDs}
\bibfield{author}{\bibinfo{person}{Marcos Serrano}, \bibinfo{person}{Barrett~M.
  Ens}, {and} \bibinfo{person}{Pourang~P. Irani}.}
  \bibinfo{year}{2014}\natexlab{}.
\newblock \showarticletitle{Exploring the Use of Hand-to-Face Input for
  Interacting with Head-Worn Displays}. In
  \bibinfo{booktitle}{\emph{Proceedings of the SIGCHI Conference on Human
  Factors in Computing Systems}} (Toronto, Ontario, Canada)
  \emph{(\bibinfo{series}{CHI '14})}. \bibinfo{publisher}{Association for
  Computing Machinery}, \bibinfo{address}{New York, NY, USA},
  \bibinfo{pages}{3181–3190}.
\newblock
\showISBNx{9781450324731}
\urldef\tempurl%
\url{https://doi.org/10.1145/2556288.2556984}
\showDOI{\tempurl}


\bibitem[\protect\citeauthoryear{Sun, Wang, Yu, Yan, Wen, and Shi}{Sun
  et~al\mbox{.}}{2017}]%
        {SunkeFloat2017}
\bibfield{author}{\bibinfo{person}{Ke Sun}, \bibinfo{person}{Yuntao Wang},
  \bibinfo{person}{Chun Yu}, \bibinfo{person}{Yukang Yan},
  \bibinfo{person}{Hongyi Wen}, {and} \bibinfo{person}{Yuanchun Shi}.}
  \bibinfo{year}{2017}\natexlab{}.
\newblock \bibinfo{booktitle}{\emph{Float: One-Handed and Touch-Free Target
  Selection on Smartwatches}}.
\newblock \bibinfo{publisher}{Association for Computing Machinery},
  \bibinfo{address}{New York, NY, USA}, \bibinfo{pages}{692–704}.
\newblock
\showISBNx{9781450346559}
\urldef\tempurl%
\url{https://doi.org/10.1145/3025453.3026027}
\showURL{%
\tempurl}


\bibitem[\protect\citeauthoryear{Vadiraj, Rao M~.V., and Ghosh}{Vadiraj
  et~al\mbox{.}}{2020}]%
        {9053124}
\bibfield{author}{\bibinfo{person}{Sanjeev~Kadagathur Vadiraj},
  \bibinfo{person}{Achuth Rao M~.V.}, {and} \bibinfo{person}{Prasanta~Kumar
  Ghosh}.} \bibinfo{year}{2020}\natexlab{}.
\newblock \showarticletitle{Automatic Identification of Speakers From Head
  Gestures in a Narration}. In \bibinfo{booktitle}{\emph{ICASSP 2020 - 2020
  IEEE International Conference on Acoustics, Speech and Signal Processing
  (ICASSP)}}. \bibinfo{pages}{6314--6318}.
\newblock
\urldef\tempurl%
\url{https://doi.org/10.1109/ICASSP40776.2020.9053124}
\showDOI{\tempurl}


\bibitem[\protect\citeauthoryear{Vu, Misra, Roy, Wei, and Lee}{Vu
  et~al\mbox{.}}{2018}]%
        {59-Smartwatch-Based}
\bibfield{author}{\bibinfo{person}{Tran~Huy Vu}, \bibinfo{person}{Archan
  Misra}, \bibinfo{person}{Quentin Roy}, \bibinfo{person}{Kenny Choo~Tsu Wei},
  {and} \bibinfo{person}{Youngki Lee}.} \bibinfo{year}{2018}\natexlab{}.
\newblock \showarticletitle{Smartwatch-Based Early Gesture Detection 8
  Trajectory Tracking for Interactive Gesture-Driven Applications}.
\newblock \bibinfo{journal}{\emph{Proc. ACM Interact. Mob. Wearable Ubiquitous
  Technol.}} \bibinfo{volume}{2}, \bibinfo{number}{1}, Article
  \bibinfo{articleno}{39} (\bibinfo{date}{March} \bibinfo{year}{2018}),
  \bibinfo{numpages}{27}~pages.
\newblock
\urldef\tempurl%
\url{https://doi.org/10.1145/3191771}
\showDOI{\tempurl}


\bibitem[\protect\citeauthoryear{Wang, Song, Lien, Poupyrev, and Hilliges}{Wang
  et~al\mbox{.}}{2016}]%
        {10.1145/2984511.2984565}
\bibfield{author}{\bibinfo{person}{Saiwen Wang}, \bibinfo{person}{Jie Song},
  \bibinfo{person}{Jaime Lien}, \bibinfo{person}{Ivan Poupyrev}, {and}
  \bibinfo{person}{Otmar Hilliges}.} \bibinfo{year}{2016}\natexlab{}.
\newblock \showarticletitle{Interacting with Soli: Exploring Fine-Grained
  Dynamic Gesture Recognition in the Radio-Frequency Spectrum}. In
  \bibinfo{booktitle}{\emph{Proceedings of the 29th Annual Symposium on User
  Interface Software and Technology}} (Tokyo, Japan)
  \emph{(\bibinfo{series}{UIST '16})}. \bibinfo{publisher}{Association for
  Computing Machinery}, \bibinfo{address}{New York, NY, USA},
  \bibinfo{pages}{851–860}.
\newblock
\showISBNx{9781450341899}
\urldef\tempurl%
\url{https://doi.org/10.1145/2984511.2984565}
\showDOI{\tempurl}


\bibitem[\protect\citeauthoryear{Wang, Ding, Chatterjee, Salemi~Parizi, Zhuang,
  Yan, Patel, and Shi}{Wang et~al\mbox{.}}{2022a}]%
        {yuntao-faceori}
\bibfield{author}{\bibinfo{person}{Yuntao Wang}, \bibinfo{person}{Jiexin Ding},
  \bibinfo{person}{Ishan Chatterjee}, \bibinfo{person}{Farshid Salemi~Parizi},
  \bibinfo{person}{Yuzhou Zhuang}, \bibinfo{person}{Yukang Yan},
  \bibinfo{person}{Shwetak Patel}, {and} \bibinfo{person}{Yuanchun Shi}.}
  \bibinfo{year}{2022}\natexlab{a}.
\newblock \showarticletitle{FaceOri: Tracking Head Position and Orientation
  Using Ultrasonic Ranging on Earphones}. In
  \bibinfo{booktitle}{\emph{Proceedings of the 2022 CHI Conference on Human
  Factors in Computing Systems}} (New Orleans, LA, USA)
  \emph{(\bibinfo{series}{CHI '22})}. \bibinfo{publisher}{Association for
  Computing Machinery}, \bibinfo{address}{New York, NY, USA}, Article
  \bibinfo{articleno}{290}, \bibinfo{numpages}{12}~pages.
\newblock
\showISBNx{9781450391573}
\urldef\tempurl%
\url{https://doi.org/10.1145/3491102.3517698}
\showDOI{\tempurl}


\bibitem[\protect\citeauthoryear{Wang, Zhang, Chakalasiya, Xu, Jiang, Li,
  Patel, and Shi}{Wang et~al\mbox{.}}{2022b}]%
        {wang-hearcough}
\bibfield{author}{\bibinfo{person}{Yuntao Wang}, \bibinfo{person}{Xiyuxing
  Zhang}, \bibinfo{person}{Jay~M. Chakalasiya}, \bibinfo{person}{Xuhai Xu},
  \bibinfo{person}{Yu Jiang}, \bibinfo{person}{Yuang Li},
  \bibinfo{person}{Shwetak Patel}, {and} \bibinfo{person}{Yuanchun Shi}.}
  \bibinfo{year}{2022}\natexlab{b}.
\newblock \showarticletitle{HearCough: Enabling continuous cough event
  detection on edge computing hearables}.
\newblock \bibinfo{journal}{\emph{Methods}}  \bibinfo{volume}{205}
  (\bibinfo{year}{2022}), \bibinfo{pages}{53--62}.
\newblock
\showISSN{1046-2023}
\urldef\tempurl%
\url{https://doi.org/10.1016/j.ymeth.2022.05.002}
\showDOI{\tempurl}


\bibitem[\protect\citeauthoryear{Wang, Hou, Jiang, Dou, Zhang, Huang, and
  Guo}{Wang et~al\mbox{.}}{2019}]%
        {Wang2019}
\bibfield{author}{\bibinfo{person}{Z Wang}, \bibinfo{person}{Y Hou},
  \bibinfo{person}{K Jiang}, \bibinfo{person}{W Dou}, \bibinfo{person}{C
  Zhang}, \bibinfo{person}{Z Huang}, {and} \bibinfo{person}{Y Guo}.}
  \bibinfo{year}{2019}\natexlab{}.
\newblock \showarticletitle{{Hand Gesture Recognition Based on Active
  Ultrasonic Sensing of Smartphone: A Survey}}.
\newblock \bibinfo{journal}{\emph{IEEE Access}}  \bibinfo{volume}{7}
  (\bibinfo{year}{2019}), \bibinfo{pages}{111897--111922}.
\newblock
\showISSN{2169-3536}
\urldef\tempurl%
\url{https://doi.org/10.1109/ACCESS.2019.2933987}
\showDOI{\tempurl}


\bibitem[\protect\citeauthoryear{Ward, Lukowicz, and Tr\"{o}ster}{Ward
  et~al\mbox{.}}{2005}]%
        {62-gesture-with-acc}
\bibfield{author}{\bibinfo{person}{Jamie~A. Ward}, \bibinfo{person}{Paul
  Lukowicz}, {and} \bibinfo{person}{Gerhard Tr\"{o}ster}.}
  \bibinfo{year}{2005}\natexlab{}.
\newblock \showarticletitle{Gesture Spotting Using Wrist Worn Microphone and
  3-Axis Accelerometer}. In \bibinfo{booktitle}{\emph{Proceedings of the 2005
  Joint Conference on Smart Objects and Ambient Intelligence: Innovative
  Context-Aware Services: Usages and Technologies}} (Grenoble, France)
  \emph{(\bibinfo{series}{sOc-EUSAI '05})}. \bibinfo{publisher}{Association for
  Computing Machinery}, \bibinfo{address}{New York, NY, USA},
  \bibinfo{pages}{99–104}.
\newblock
\showISBNx{1595933042}
\urldef\tempurl%
\url{https://doi.org/10.1145/1107548.1107578}
\showDOI{\tempurl}


\bibitem[\protect\citeauthoryear{Wen, Ramos~Rojas, and Dey}{Wen
  et~al\mbox{.}}{2016}]%
        {Serendipity}
\bibfield{author}{\bibinfo{person}{Hongyi Wen}, \bibinfo{person}{Julian
  Ramos~Rojas}, {and} \bibinfo{person}{Anind~K. Dey}.}
  \bibinfo{year}{2016}\natexlab{}.
\newblock \bibinfo{booktitle}{\emph{Serendipity: Finger Gesture Recognition
  Using an Off-the-Shelf Smartwatch}}.
\newblock \bibinfo{publisher}{Association for Computing Machinery},
  \bibinfo{address}{New York, NY, USA}, \bibinfo{pages}{3847–3851}.
\newblock
\showISBNx{9781450333627}
\urldef\tempurl%
\url{https://doi.org/10.1145/2858036.2858466}
\showURL{%
\tempurl}


\bibitem[\protect\citeauthoryear{Weng, Yu, Shi, Zhao, Yan, and Shi}{Weng
  et~al\mbox{.}}{2021}]%
        {46-facesight}
\bibfield{author}{\bibinfo{person}{Yueting Weng}, \bibinfo{person}{Chun Yu},
  \bibinfo{person}{Yingtian Shi}, \bibinfo{person}{Yuhang Zhao},
  \bibinfo{person}{Yukang Yan}, {and} \bibinfo{person}{Yuanchun Shi}.}
  \bibinfo{year}{2021}\natexlab{}.
\newblock \showarticletitle{FaceSight: Enabling Hand-to-Face Gesture
  Interaction on AR Glasses with a Downward-Facing Camera Vision}. In
  \bibinfo{booktitle}{\emph{Proceedings of the 2021 CHI Conference on Human
  Factors in Computing Systems}}. \bibinfo{pages}{1--14}.
\newblock


\bibitem[\protect\citeauthoryear{Wu, Harrison, Bigham, and Laput}{Wu
  et~al\mbox{.}}{2020}]%
        {Wu2020}
\bibfield{author}{\bibinfo{person}{Jason Wu}, \bibinfo{person}{Chris Harrison},
  \bibinfo{person}{Jeffrey~P. Bigham}, {and} \bibinfo{person}{Gierad Laput}.}
  \bibinfo{year}{2020}\natexlab{}.
\newblock \showarticletitle{{Automated Class Discovery and One-Shot
  Interactions for Acoustic Activity Recognition}}.
\newblock \bibinfo{journal}{\emph{Conference on Human Factors in Computing
  Systems - Proceedings}} (\bibinfo{year}{2020}).
\newblock
\showISBNx{9781450367080}
\urldef\tempurl%
\url{https://doi.org/10.1145/3313831.3376875}
\showDOI{\tempurl}


\bibitem[\protect\citeauthoryear{Xiao, Lew, Marsanico, Hariharan, Hudson, and
  Harrison}{Xiao et~al\mbox{.}}{2014}]%
        {Xiao-2014-Toffee}
\bibfield{author}{\bibinfo{person}{Robert Xiao}, \bibinfo{person}{Greg Lew},
  \bibinfo{person}{James Marsanico}, \bibinfo{person}{Divya Hariharan},
  \bibinfo{person}{Scott Hudson}, {and} \bibinfo{person}{Chris Harrison}.}
  \bibinfo{year}{2014}\natexlab{}.
\newblock \showarticletitle{{Toffee: Enabling Ad Hoc, Around-device Interaction
  with Acoustic Time-of-arrival Correlation}}. In
  \bibinfo{booktitle}{\emph{Proceedings of the 16th International Conference on
  Human-computer Interaction with Mobile Devices {\&}{\#}38; Services}}
  \emph{(\bibinfo{series}{MobileHCI '14})}. \bibinfo{publisher}{ACM},
  \bibinfo{address}{New York, NY, USA}, \bibinfo{pages}{67--76}.
\newblock
\showISBNx{978-1-4503-3004-6}
\urldef\tempurl%
\url{https://doi.org/10.1145/2628363.2628383}
\showDOI{\tempurl}


\bibitem[\protect\citeauthoryear{Xu, Shi, Yi, Liu, Yan, Shi, Mariakakis,
  Mankoff, and Dey}{Xu et~al\mbox{.}}{2020}]%
        {earbuddy}
\bibfield{author}{\bibinfo{person}{Xuhai Xu}, \bibinfo{person}{Haitian Shi},
  \bibinfo{person}{Xin Yi}, \bibinfo{person}{WenJia Liu},
  \bibinfo{person}{Yukang Yan}, \bibinfo{person}{Yuanchun Shi},
  \bibinfo{person}{Alex Mariakakis}, \bibinfo{person}{Jennifer Mankoff}, {and}
  \bibinfo{person}{Anind~K. Dey}.} \bibinfo{year}{2020}\natexlab{}.
\newblock \bibinfo{booktitle}{\emph{EarBuddy: Enabling On-Face Interaction via
  Wireless Earbuds}}.
\newblock \bibinfo{publisher}{Association for Computing Machinery},
  \bibinfo{address}{New York, NY, USA}, \bibinfo{pages}{1–14}.
\newblock
\showISBNx{9781450367080}
\urldef\tempurl%
\url{https://doi.org/10.1145/3313831.3376836}
\showURL{%
\tempurl}


\bibitem[\protect\citeauthoryear{Yamashita, Kikuchi, Masai, Sugimoto, Thomas,
  and Sugiura}{Yamashita et~al\mbox{.}}{2017}]%
        {cheek}
\bibfield{author}{\bibinfo{person}{Koki Yamashita}, \bibinfo{person}{Takashi
  Kikuchi}, \bibinfo{person}{Katsutoshi Masai}, \bibinfo{person}{Maki
  Sugimoto}, \bibinfo{person}{Bruce~H. Thomas}, {and} \bibinfo{person}{Yuta
  Sugiura}.} \bibinfo{year}{2017}\natexlab{}.
\newblock \showarticletitle{CheekInput: Turning Your Cheek into an Input
  Surface by Embedded Optical Sensors on a Head-Mounted Display}. In
  \bibinfo{booktitle}{\emph{Proceedings of the 23rd ACM Symposium on Virtual
  Reality Software and Technology}} (Gothenburg, Sweden)
  \emph{(\bibinfo{series}{VRST '17})}. \bibinfo{publisher}{Association for
  Computing Machinery}, \bibinfo{address}{New York, NY, USA}, Article
  \bibinfo{articleno}{19}, \bibinfo{numpages}{8}~pages.
\newblock
\showISBNx{9781450355483}
\urldef\tempurl%
\url{https://doi.org/10.1145/3139131.3139146}
\showDOI{\tempurl}


\bibitem[\protect\citeauthoryear{Yan, Yu, Shi, and Xie}{Yan
  et~al\mbox{.}}{2019}]%
        {Yan-UIST-2019}
\bibfield{author}{\bibinfo{person}{Yukang Yan}, \bibinfo{person}{Chun Yu},
  \bibinfo{person}{Yingtian Shi}, {and} \bibinfo{person}{Minxing Xie}.}
  \bibinfo{year}{2019}\natexlab{}.
\newblock \showarticletitle{{PrivateTalk: Activating voice input with
  hand-on-mouth gesture detected by bluetooth earphones}}. In
  \bibinfo{booktitle}{\emph{UIST 2019 - Proceedings of the 32nd Annual ACM
  Symposium on User Interface Software and Technology}}.
  \bibinfo{pages}{1013--1020}.
\newblock
\showISBNx{9781450368162}
\urldef\tempurl%
\url{https://doi.org/10.1145/3332165.3347950}
\showDOI{\tempurl}


\bibitem[\protect\citeauthoryear{Yan, Yu, Zheng, Tang, Xu, and Shi}{Yan
  et~al\mbox{.}}{2020}]%
        {10.1145/3313831.3376810}
\bibfield{author}{\bibinfo{person}{Yukang Yan}, \bibinfo{person}{Chun Yu},
  \bibinfo{person}{Wengrui Zheng}, \bibinfo{person}{Ruining Tang},
  \bibinfo{person}{Xuhai Xu}, {and} \bibinfo{person}{Yuanchun Shi}.}
  \bibinfo{year}{2020}\natexlab{}.
\newblock \bibinfo{booktitle}{\emph{FrownOnError: Interrupting Responses from
  Smart Speakers by Facial Expressions}}.
\newblock \bibinfo{publisher}{Association for Computing Machinery},
  \bibinfo{address}{New York, NY, USA}, \bibinfo{pages}{1–14}.
\newblock
\showISBNx{9781450367080}
\urldef\tempurl%
\url{https://doi.org/10.1145/3313831.3376810}
\showURL{%
\tempurl}


\bibitem[\protect\citeauthoryear{Yang, Yu, Zheng, and Shi}{Yang
  et~al\mbox{.}}{2019}]%
        {10.1145/3351276}
\bibfield{author}{\bibinfo{person}{Zhican Yang}, \bibinfo{person}{Chun Yu},
  \bibinfo{person}{Fengshi Zheng}, {and} \bibinfo{person}{Yuanchun Shi}.}
  \bibinfo{year}{2019}\natexlab{}.
\newblock \showarticletitle{ProxiTalk: Activate Speech Input by Bringing
  Smartphone to the Mouth}.
\newblock \bibinfo{journal}{\emph{Proc. ACM Interact. Mob. Wearable Ubiquitous
  Technol.}} \bibinfo{volume}{3}, \bibinfo{number}{3}, Article
  \bibinfo{articleno}{118} (\bibinfo{date}{sep} \bibinfo{year}{2019}),
  \bibinfo{numpages}{25}~pages.
\newblock
\urldef\tempurl%
\url{https://doi.org/10.1145/3351276}
\showDOI{\tempurl}


\bibitem[\protect\citeauthoryear{Yu, Wei, Vachher, Qin, Liang, Weng, Gu, and
  Shi}{Yu et~al\mbox{.}}{2019}]%
        {55-handsee}
\bibfield{author}{\bibinfo{person}{Chun Yu}, \bibinfo{person}{Xiaoying Wei},
  \bibinfo{person}{Shubh Vachher}, \bibinfo{person}{Yue Qin},
  \bibinfo{person}{Chen Liang}, \bibinfo{person}{Yueting Weng},
  \bibinfo{person}{Yizheng Gu}, {and} \bibinfo{person}{Yuanchun Shi}.}
  \bibinfo{year}{2019}\natexlab{}.
\newblock \bibinfo{booktitle}{\emph{HandSee: Enabling Full Hand Interaction on
  Smartphone with Front Camera-Based Stereo Vision}}.
\newblock \bibinfo{publisher}{Association for Computing Machinery},
  \bibinfo{address}{New York, NY, USA}, \bibinfo{pages}{1–13}.
\newblock
\showISBNx{9781450359702}
\urldef\tempurl%
\url{https://doi.org/10.1145/3290605.3300935}
\showURL{%
\tempurl}


\bibitem[\protect\citeauthoryear{Yun, Chen, Zheng, Qiu, and Mao}{Yun
  et~al\mbox{.}}{2017}]%
        {Yun2017}
\bibfield{author}{\bibinfo{person}{Sangki Yun}, \bibinfo{person}{Yi~Chao Chen},
  \bibinfo{person}{Huihunag Zheng}, \bibinfo{person}{Lili Qiu}, {and}
  \bibinfo{person}{Wenguang Mao}.} \bibinfo{year}{2017}\natexlab{}.
\newblock \showarticletitle{{Strata: Fine-grained acoustic-based device-free
  tracking}}.
\newblock \bibinfo{journal}{\emph{MobiSys 2017 - Proceedings of the 15th Annual
  International Conference on Mobile Systems, Applications, and Services}}
  (\bibinfo{year}{2017}), \bibinfo{pages}{15--28}.
\newblock
\showISBNx{9781450349284}
\urldef\tempurl%
\url{https://doi.org/10.1145/3081333.3081356}
\showDOI{\tempurl}


\bibitem[\protect\citeauthoryear{Zhang, Zhou, Lin, and Sun}{Zhang
  et~al\mbox{.}}{2018}]%
        {zhang2018shufflenet}
\bibfield{author}{\bibinfo{person}{Xiangyu Zhang}, \bibinfo{person}{Xinyu
  Zhou}, \bibinfo{person}{Mengxiao Lin}, {and} \bibinfo{person}{Jian Sun}.}
  \bibinfo{year}{2018}\natexlab{}.
\newblock \showarticletitle{Shufflenet: An extremely efficient convolutional
  neural network for mobile devices}. In \bibinfo{booktitle}{\emph{Proceedings
  of the IEEE conference on computer vision and pattern recognition}}.
  \bibinfo{pages}{6848--6856}.
\newblock


\bibitem[\protect\citeauthoryear{Zhang, Huang, Yang, Wang, Chen, You, Huang,
  Xue, and Yu}{Zhang et~al\mbox{.}}{2020}]%
        {Zhang2020}
\bibfield{author}{\bibinfo{person}{Yongzhao Zhang}, \bibinfo{person}{Wei~Hsiang
  Huang}, \bibinfo{person}{Chih~Yun Yang}, \bibinfo{person}{Wen~Ping Wang},
  \bibinfo{person}{Yi~Chao Chen}, \bibinfo{person}{Chuang~Wen You},
  \bibinfo{person}{Da~Yuan Huang}, \bibinfo{person}{Guangtao Xue}, {and}
  \bibinfo{person}{Jiadi Yu}.} \bibinfo{year}{2020}\natexlab{}.
\newblock \showarticletitle{{Endophasia: Utilizing acoustic-based imaging for
  issuing contact-free silent speech commands}}.
\newblock \bibinfo{journal}{\emph{Proceedings of the ACM on Interactive,
  Mobile, Wearable and Ubiquitous Technologies}} \bibinfo{volume}{4},
  \bibinfo{number}{1} (\bibinfo{year}{2020}).
\newblock
\showISSN{24749567}
\urldef\tempurl%
\url{https://doi.org/10.1145/3381008}
\showDOI{\tempurl}


\bibitem[\protect\citeauthoryear{Zhuang, Wang, Yan, Xu, and Shi}{Zhuang
  et~al\mbox{.}}{2021}]%
        {Yuzhou-reflectrack}
\bibfield{author}{\bibinfo{person}{Yuzhou Zhuang}, \bibinfo{person}{Yuntao
  Wang}, \bibinfo{person}{Yukang Yan}, \bibinfo{person}{Xuhai Xu}, {and}
  \bibinfo{person}{Yuanchun Shi}.} \bibinfo{year}{2021}\natexlab{}.
\newblock \showarticletitle{ReflecTrack: Enabling 3D Acoustic Position Tracking
  Using Commodity Dual-Microphone Smartphones}. In
  \bibinfo{booktitle}{\emph{The 34th Annual ACM Symposium on User Interface
  Software and Technology}} (Virtual Event, USA) \emph{(\bibinfo{series}{UIST
  '21})}. \bibinfo{publisher}{Association for Computing Machinery},
  \bibinfo{address}{New York, NY, USA}, \bibinfo{pages}{1050–1062}.
\newblock
\showISBNx{9781450386357}
\urldef\tempurl%
\url{https://doi.org/10.1145/3472749.3474805}
\showDOI{\tempurl}


\end{thebibliography}

\appendix

\end{document}